\documentclass[a4paper]{article}
\PassOptionsToPackage{bookmarks=false}{hyperref}
\usepackage{ISCSLP2021,subfigure,multirow,array,cite}
\usepackage[backref]{hyperref}
\hypersetup{
colorlinks=true,
linkcolor=black,
citecolor=black,
urlcolor=black
}

\usepackage{ISCSLP2021}

\usepackage{cite}
\usepackage{amssymb,amsfonts}
\usepackage{algorithmic}
\usepackage{float}
\usepackage{stfloats}
\usepackage{subfigure}
\usepackage{amssymb,amsmath,bm}
\usepackage{soul}
\usepackage{tabularx}
\usepackage{textcomp}
\usepackage{multirow}
\usepackage{caption}
\usepackage{booktabs}
\usepackage{makecell}
\usepackage{url}

\title{Controllable Emotion Transfer For End-to-End Speech Synthesis}
\name{Tao Li, Shan Yang, Liumeng Xue, Lei Xie$^*$}

\address{Audio, Speech and Language Processing Group (ASLP@NPU), School of Computer Science, \\Northwestern Polytechnical University, Xi'an, China}
\email{taoli@npu-aslp.org, \{syang, lmxue, lxie\}@nwpu-aslp.org}

\begin{document}
%\ninept

\maketitle

\begin{abstract}
Emotion embedding space learned from references is a straightforward approach for emotion transfer in encoder-decoder structured emotional text to speech (TTS) systems. However, the transferred emotion in the synthetic speech is not accurate and expressive enough with emotion category confusions. Moreover, it is hard to select an appropriate reference to deliver desired emotion strength. To solve these problems, we propose a novel approach based on Tacotron. First, we plug two emotion classifiers -- one after the reference encoder, one after the decoder output -- to enhance the emotion-discriminative ability of the emotion embedding and the predicted mel-spectrum. Second, we adopt style loss to measure the difference between the generated and reference mel-spectrum. The emotion strength in the synthetic speech can be controlled by adjusting the value of the emotion embedding as the emotion embedding can be viewed as the feature map of the mel-spectrum. Experiments on emotion transfer and strength control have shown that the synthetic speech of the proposed method is more accurate and expressive with less emotion category confusions and the control of emotion strength is more salient to listeners.

\end{abstract}
\renewcommand{\thefootnote}{\fnsymbol{footnote}}
\footnotetext{* Lei Xie is the corresponding author. This work was supported by the National Key Research and Development Program of China under Grant 2017YFB1002102.}

\vspace{-5pt}
\section{Introduction}
\label{sec:intro}

The naturalness of speech synthesis has been dramatically advanced with the proliferation of sequence-to-sequence (seq2seq) based neural approaches~\cite{wang2017tacotron,shen2018natural,Ren2019FastSpeechFR,yu2019durian,Ling2015DeepLF}. As natural sound can be reasonably produced by current seq2seq-based models learned from a typical corpus with consistently neutral speaking style, there has been increasing interest in how to deliver expressive speech with these seq2seq models~\cite{Skerry2018Towards,Wang2018Style, Zhang2019LearningLR,Wang2017Uncovering,Bian2019Multi}. Human speech is expressive in nature and delivering accurate and controllable expressive speech from text is highly desired with substantial applications in human-computer interaction and audio content generation. Proper expression rendering affects overall speech perception, which is important for applications such as audiobooks and newsreaders. In particular, emotional speech synthesis, which focuses on emotion expression rendering, has drawn much attention recently~\cite{Ohtani2015EmotionalTI,Inoue2017AnIT,Choi2019MultispeakerEA,Johnson2016Perceptual,Se2020Emotional}. The emotional expressions are directly affected by the speaker's intentions, leading to speech with different emotion categories such as happy, angry, sad and fear. This paper addresses the emotional speech synthesis problem under the seq2seq-based paradigm.

%Compared to the conventional speech synthesis~\cite{Wang2017LearningUR,LorenzoTrueba2018InvestigatingDR,Henter2018DeepEM,Bulut2002ExpressiveSS}, neural attention-based sequence-to-sequence (seq2seq) approaches has significantly improved the performance of neutral speech synthesis~\cite{Oord2016WaveNet,Wang2017Tacotron,Sotelo2017Char2WavES,Arik2017Deep,Gibiansky2017DeepV2}. To achieve more natural and human-like synthesized speech, emotional speech synthesis has achieved growing interests~\cite{Ohtani2015EmotionalTI,Inoue2017AnIT,Choi2019MultispeakerEA,Johnson2016Perceptual}.

There are two key problems that need to be addressed for emotional speech synthesis to achieve ideally human-parity performance. First, delivering the emotion \textit{properly and accurately} through synthesized speech. Specifically, the emotion conveyed by the synthetic speech should be perceived easily by the listeners without confusions. Second, controlling the emotion delivery in a \textit{flexible} way. The emotion expressions embedded in human speech are subtle with different \textit{strengths}. Thus we desire an emotional TTS to flexibly deliver emotional speech with preferred strength. For example, we should manage to synthesize `very happy' and `a little bit happy' through synthetic speech.

There has been a long history for tackling the first problem, which can trace back to the era of conventional Hidden Markov Model (HMM) based statistical parametric speech synthesis~\cite{Yamagishi2005AcousticMO,Qin2006HMMBasedES,LorenzoTrueba2015EmotionTT,An2017EmotionalSP}. A straight-forward way to conduct emotional speech synthesis is to use categorized emotional data to train a model~\cite{Lee2017Emotional}. When emotional data is limited with only a few samples, model adaptation~\cite{Ohtani2015Emotional,Inoue2017An,Yang2016OnTT} is often adopted on the average or neutral voice model that has been trained beforehand using a sizable set of data. With the wide use of deep neural networks~\cite{Inoue2017AnIT,Ling2015DeepLF,Xue2018ACO}, besides the above adaptation method, emotional speech synthesis has evolved to multiple solutions, such as code/embedding-based~\cite{Choi2019MultispeakerEA,Wu2018RapidSA,Li2018EmphaticSG} input and multi-head network~\cite{Wang2018Style}. However, these studies only can learn an \textit{averaged} emotion distribution over the training data, lacking the ability for fine-grained control. Another raised problem is \textit{emotion confusion}, i.e., synthetic emotional speech samples always have confusion over different emotion categories.

To the best of our knowledge, there are few studies addressing the second problem due to the difficulties on how to label and quantize emotion strength given an emotion speech corpus. A recent study has managed to approach this problem by learning a ranking function on emotional speech samples using relative attributes in an unsupervised manner~\cite{Zhu2019ControllingES,Ferrari2008Learning}. Thus the learned ranking function can designate each sample a strength that is subsequently used as a label to train a Tacotron-based TTS model.

With the fast development of the seq2seq modeling architecture, particularly the Tacotron family~\cite{wang2017tacotron,shen2018natural}, \textit{reference-based style transfer} has emerged as another solution with great potential to solving the two problems simultaneously. By learning a style embedding space through expressive samples in an unsupervised manner and conditioning Tacotron on it, synthetic audio that matches the prosody of the reference audio can be generated even when the reference and synthesis speakers are different. Following the principle of `\textit{say it like this}'~\cite{Skerry2018Towards}, there has evolved a plenty of work in this direction lately~\cite{Wu2018FeatureBA}, such as Global Style Tokens (GST)~\cite{Wang2018Style,Se2020Emotional}, Variational Autoencoder (VAE)~\cite{Zhang2019LearningLR,Kingma2014AutoEncodingVB} and their variants and updates~\cite{Ma2019NeuralTS, 2019Multi}. However, the transferred emotion in the synthetic audio is often over-averaged and it is hard to select a proper reference to deliver the desired emotion strength as well.

To solve the two problems -- deliver emotion accurately and control emotion strength flexibly, we propose a reference-based emotion speech synthesis approach based on the Tacotron framework. Specifically, similar to ~\cite{Skerry2018Towards, Wang2018Style}, we adopt a reference encoder to learn an emotion embedding space and Tacotron is conditioned on the emotion embedding for emotion transfer. But differently, in order to transfer emotion accurately and expressively, we use an emotion classifier connected to the reference encoder to enhance the emotion-discriminative ability of the emotion space. Moreover, we use another reference encoder with an emotion classifier after the decoder output to further strengthen the emotion-discriminative ability of the predicted mel-spectrum. Importantly, we further use style loss~\cite{GatysA,Johnson2016Perceptual} to measure the style difference between the generated and reference mel-spectrum. In detail, we calculate the Gram matrices of the above embeddings and then minimize the L2 distance between them. The emotion strength in the synthetic speech can be controlled easily by adjusting the value of the emotion embedding. Our model is thus learned with the integration of four losses: the basic Tacotron MSE loss, two emotion classification losses and the style loss. Experiments on emotion transfer and strength control have validated the effectiveness of our approach: synthetic speech is more accurate and expressive with less emotion category confusions; the control of emotion strength is more salient to subjective listeners.

%In this paper, we propose a controllable style embedding to learn the  strength of imitation for controllable emotion transfer. Specifically, we propose an emotion embedding space learning network to learn a fixed-length emotion embedding from a reference audio, which contains information of both emotion category and  strength. To enhance the category representation, we utilize an auxiliary emotion classifier to force the module to distinguish the emotion categories. As for the strength of emotion transfer, we firstly adopt another emotion embedding space learning network to learn the emotional representations from the synthesized speech. And then we minimize the style distance between the emotion embedding learned from reference speech and synthesized speech with the help of style loss~\cite{GatysA,Johnson2016Perceptual}. In details, we calculate the Gram matrices of the above embeddings and then minimize the $L2$ distance between them. With the proposed emotion embedding, experimental results show that the our model can maintain the category and flexibly control the strength of transfer given an emotional reference audio.

\begin{figure*}[htb]	
	\centering
	\includegraphics[width=16.5cm]{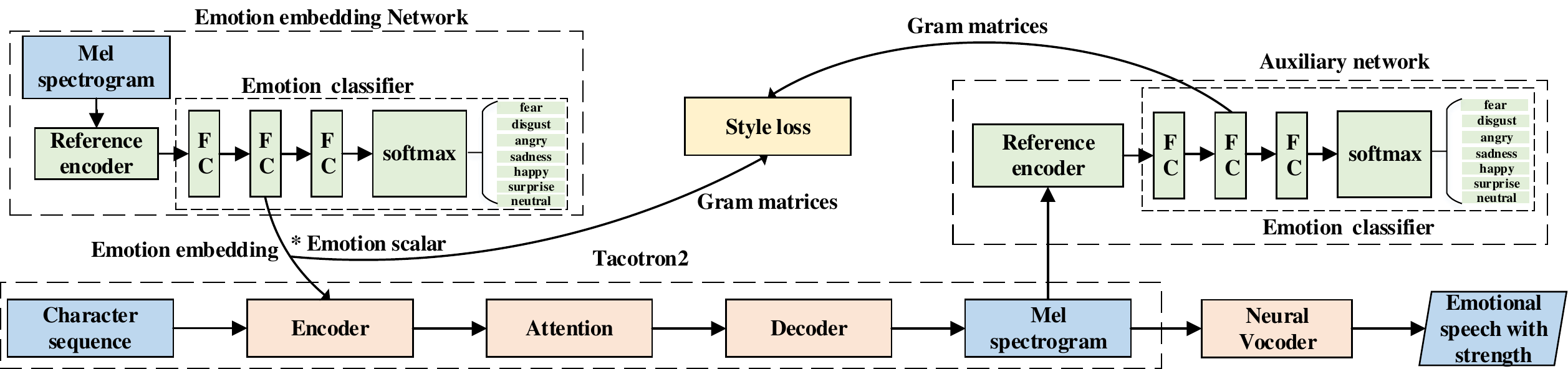}
    \captionsetup{belowskip=-18pt}
	\caption{The architecture of the proposed model }
	\label{fig:frame_diagram}
\end{figure*}

\vspace{-5pt}

\section{The proposed model}
The proposed model is shown in Figure~\ref{fig:frame_diagram}, which is built on the modified Tacotron2~\cite{shen2018natural} with an emotion embedding learning network, an auxiliary learning network and a specifically designed style loss.

\vspace{-5pt}
 \subsection{The modified Tacotron2}
\label{sec:pagestyle}
We use a slightly modified version of Tacotron2 which empirically shows better performance. First, we convert the input Chinese text into a character sequence. The encoder is composed of a pre-net of two fully connected layers and a CBHG~\cite{Lee2017Fully} (1-D convolution bank + highway network + bidirectional GRU~\cite{Cho2014LearningPR}) module. The CBHG module converts the pre-net output into the final encoder representation, followed by GMM attention. The decoder is an autoregressive recurrent neural network (RNN) in which a stack of GRUs with vertical residual connections generating attention queries at each decoder time step. Finally, we use a CBHG-based post-net to transform the mel-scale spectrogram into a linear spectrogram for reconstructing waveform by a multi-band WaveRNN~\cite{yu2019durian}.

\vspace{-5pt}
 \subsection{Emotion embedding network}
\vspace{-2pt}
The emotion embedding network is shown on the top left of Figure~\ref{fig:frame_diagram}, which is comprised of a reference encoder and an emotion classifier. We design this network to construct an emotion embedding space, which learns from reference audio samples and performs emotion transfer during inference. It also can be adjusted via a continuous scalar to control the strength of emotion transferred.

\textbf{The reference encoder.} The reference encoder follows the structure proposed by Skerry-Ryan \textit{et al.}~\cite{Skerry2018Towards}, which consists of six 2D convolutional layers and a GRU~\cite{Cho2014LearningPR} layer, and the last GRU state passes through a fully connected layer (FC) to generate a 128-dimensional reference embedding.

\textbf{Emotion classifier.} Different from~\cite{Skerry2018Towards}, we plug an emotion classifier to the reference encoder, which aims to learn more discriminative emotion embedding that can better distinguish different emotion types. In detail, the classifier has a 128-unit input layer and two 256-unit fully connected (FC) layers, both with ReLu activation. The final softmax layer outputs the probability of seven emotion types, i.e., neutral, happy, surprise, angry, disgust, fear and sad. We use the second hidden layer output as the emotion embedding and the Tacotron2 encoder output takes it as a condition.

\vspace{-5pt}
 \subsection{Auxiliary network}

We use another auxiliary emotion classifier along with the decoder to further make the predicted mel-spectrogram more discriminative to emotion types. As shown on the top right of Figure~\ref{fig:frame_diagram}, the structure of this classification network is same as that of the emotion embedding network, but the input is the predicted mel-spectrogram from the decoder. The second hidden layer output is also used as the emotion representation of the synthesized speech.

\vspace{-5pt}
\subsection{Style loss}
\begin{figure*}[t]
	\centering
	\begin{minipage}{\linewidth}
		\begin{minipage}{0.015\linewidth}
			\centerline{\includegraphics[width=\textwidth]{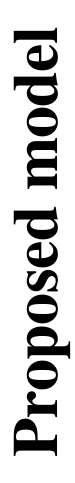}}
		\end{minipage}
		\hfill		
		\begin{minipage}{0.158\linewidth}
			\centerline{\includegraphics[width=\textwidth]{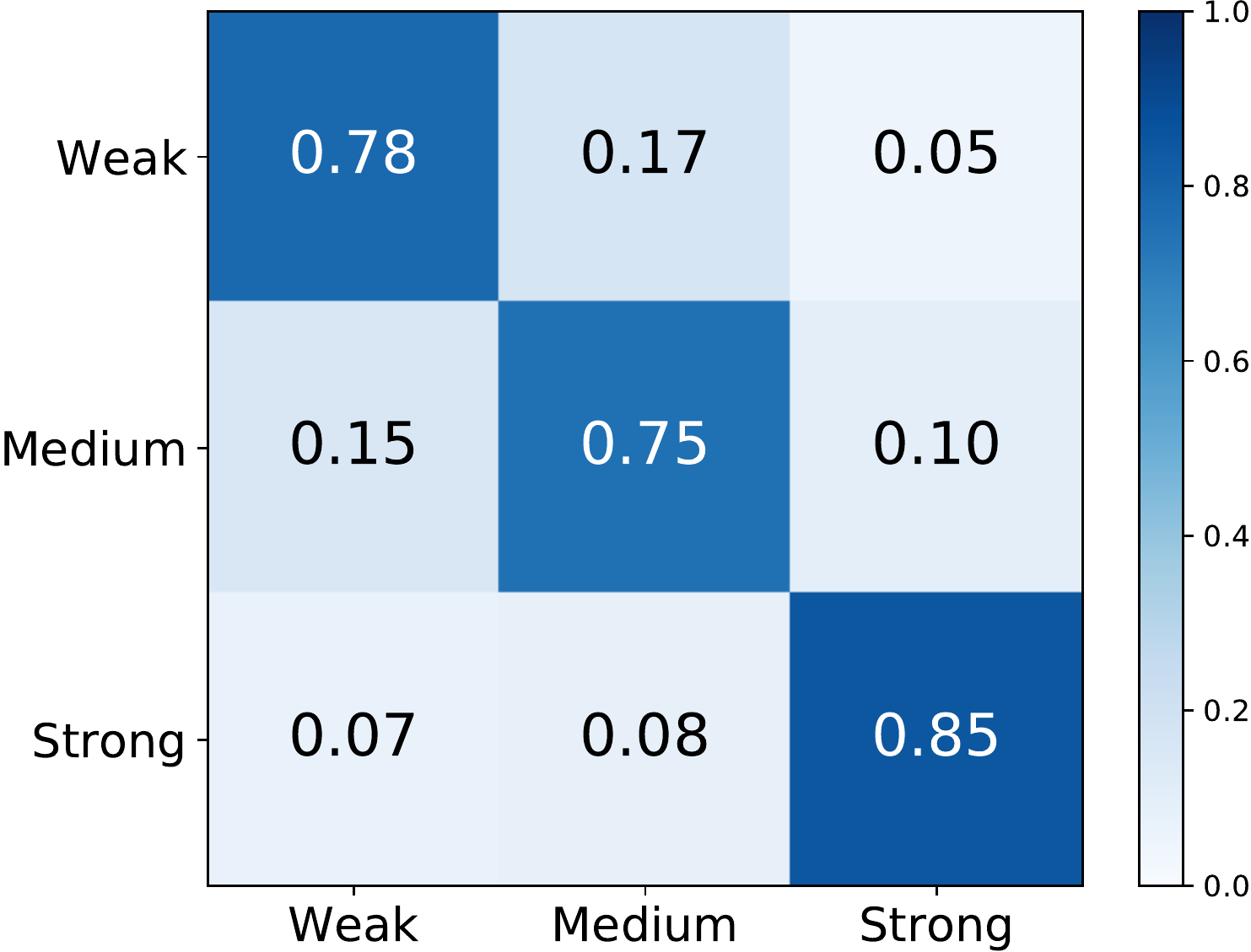}}
			%\centerline{Surprise}
		\end{minipage}
		\hfill
		\begin{minipage}{0.158\linewidth}
			\centerline{\includegraphics[width=\textwidth]{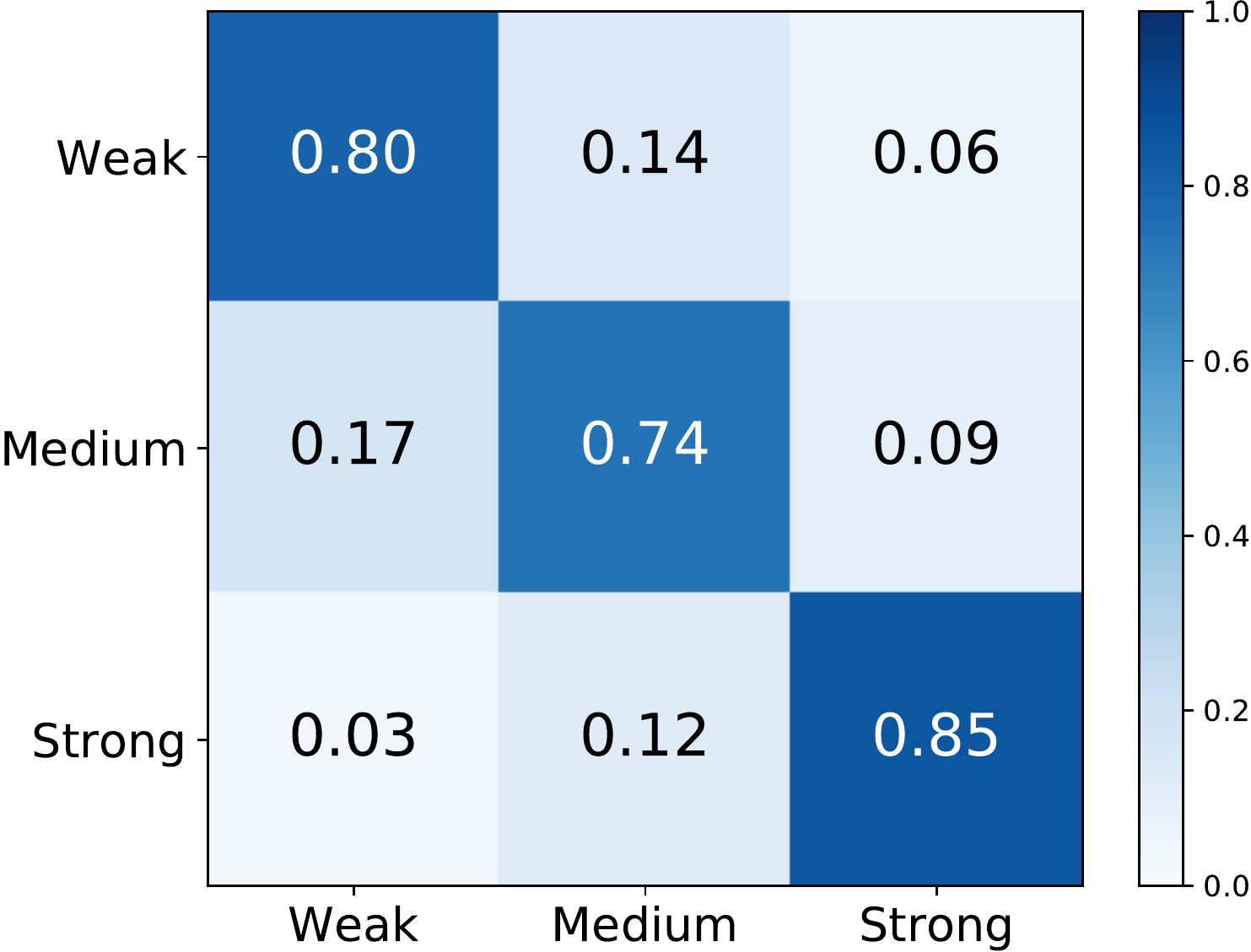}}
			%\centerline{Happiness}
		\end{minipage}
		\hfill
		\begin{minipage}{0.158\linewidth}
			\centerline{\includegraphics[width=\textwidth]{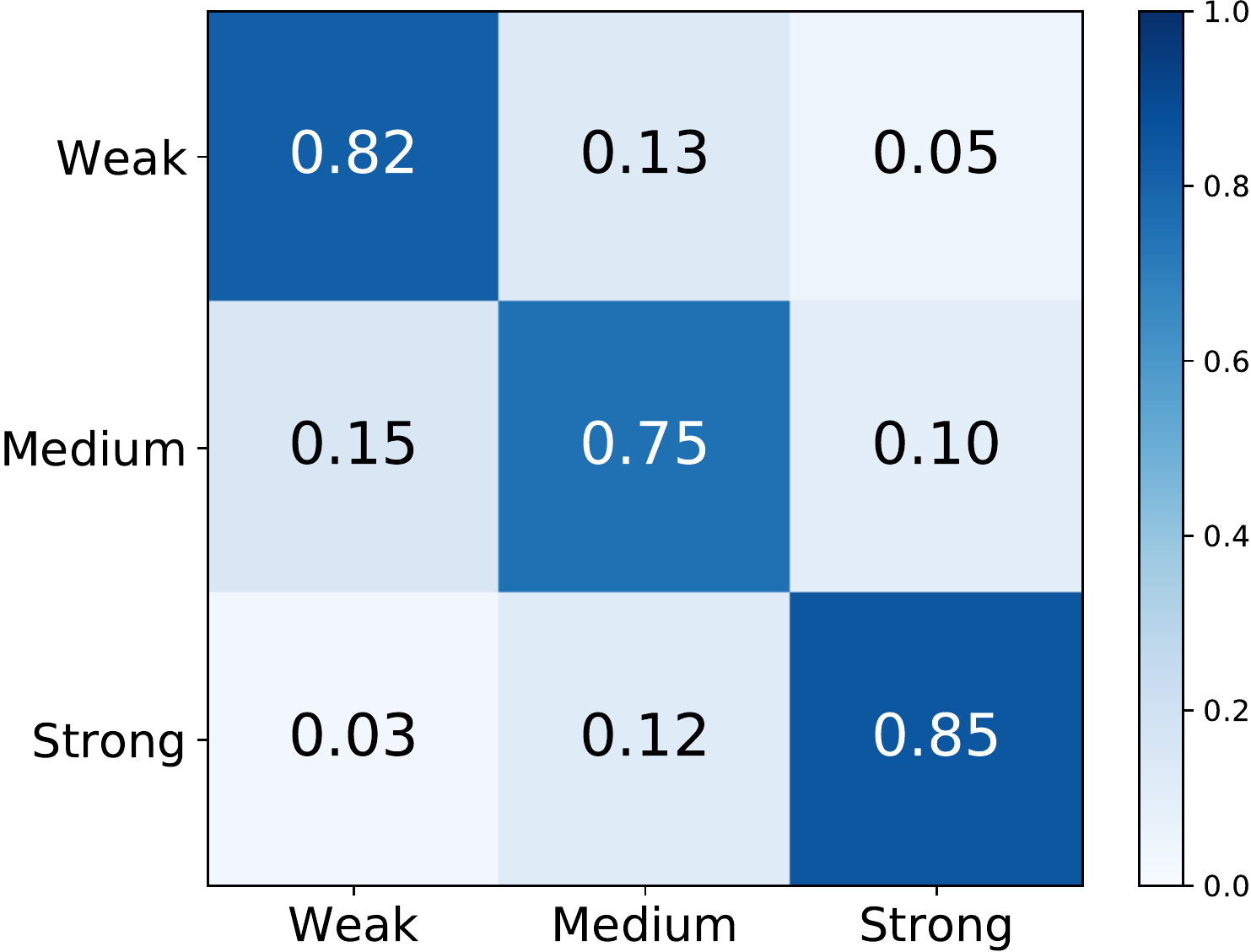}}
			%\centerline{Sadness}
		\end{minipage}
		\hfill
		\begin{minipage}{0.158\linewidth}
			\centerline{\includegraphics[width=\textwidth]{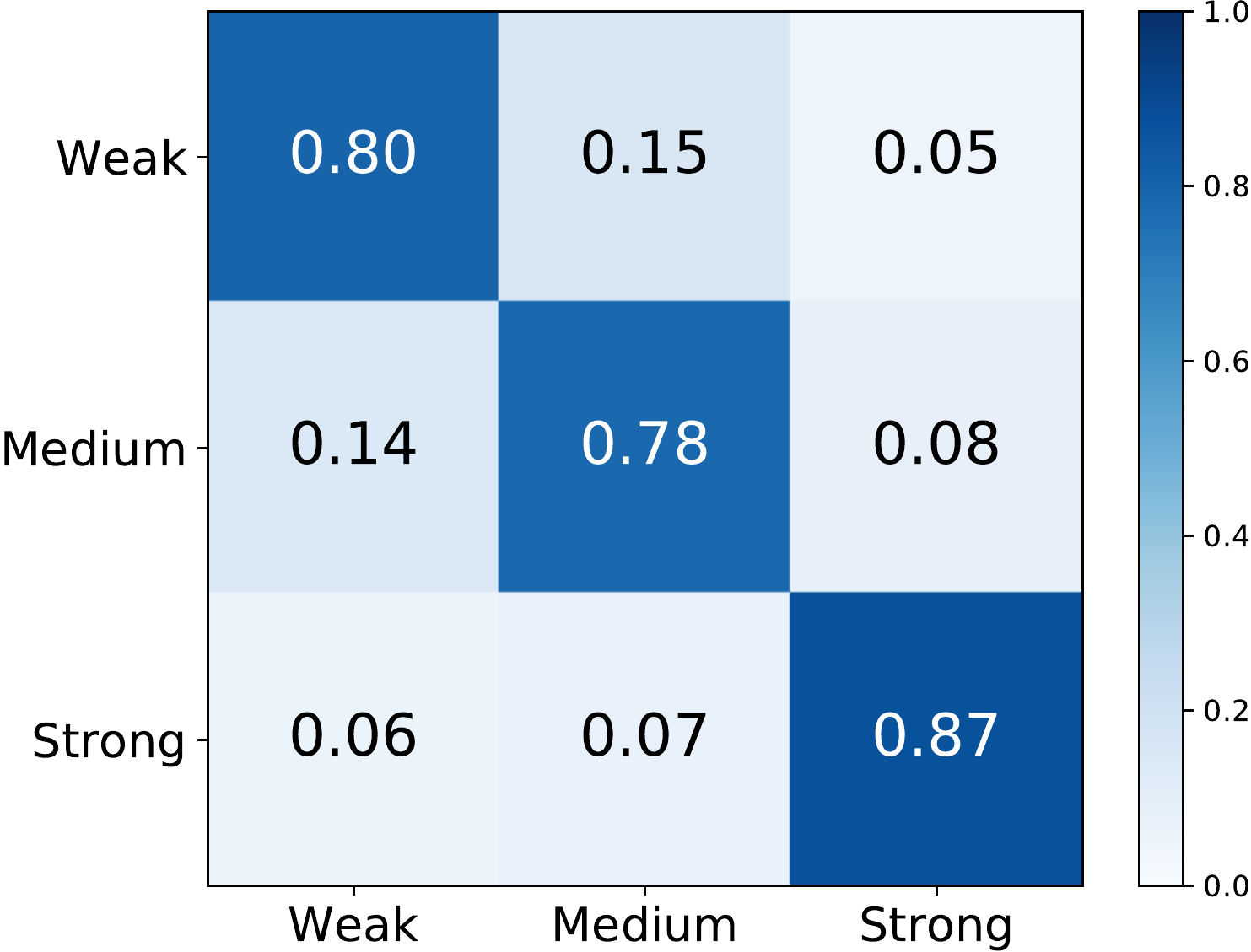}}
			%\centerline{Anger}
		\end{minipage}
		\hfill
		\begin{minipage}{0.158\linewidth}
			\centerline{\includegraphics[width=\textwidth]{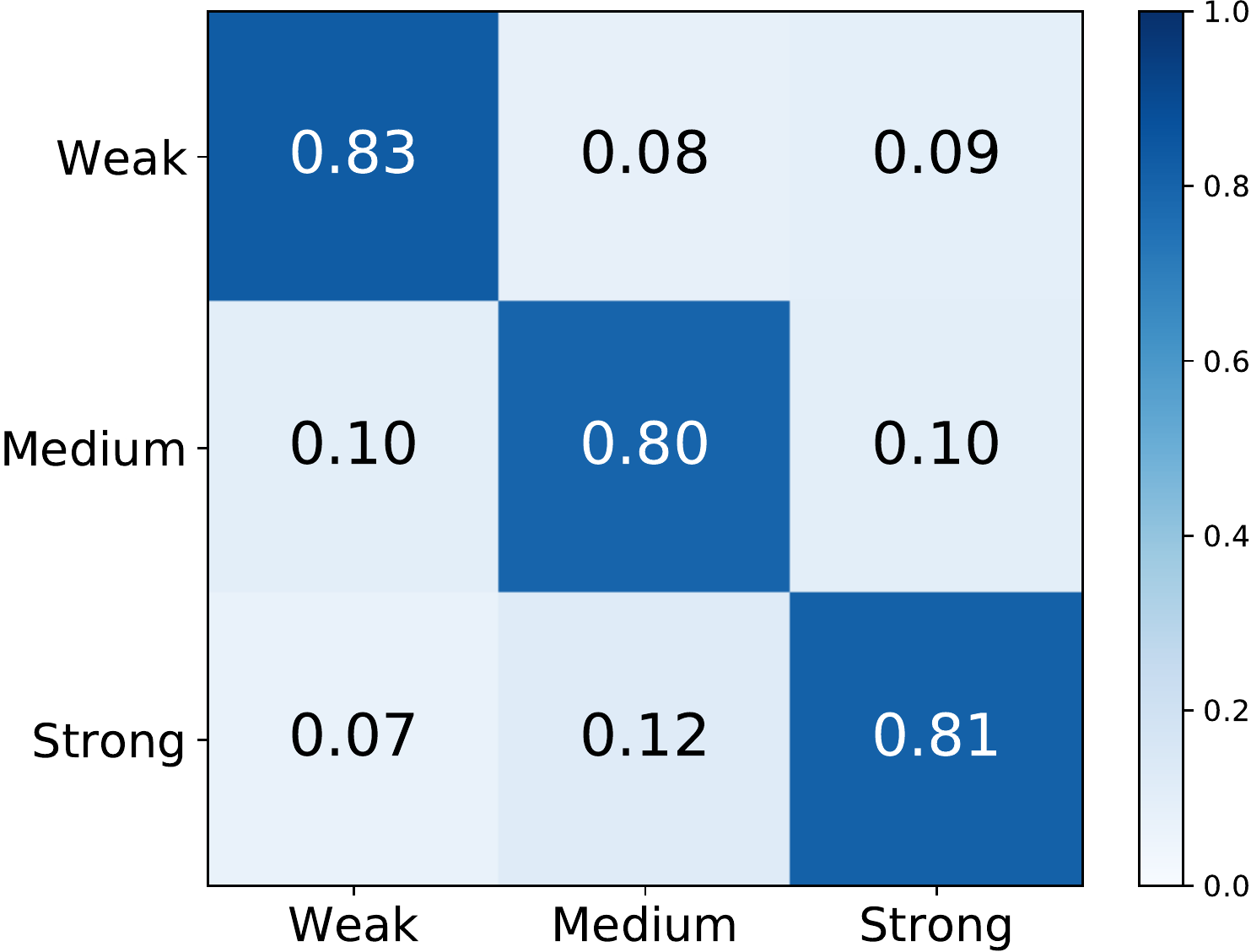}}
			%\centerline{Disgust}
		\end{minipage}
		\hfill
		\begin{minipage}{0.158\linewidth}
			\centerline{\includegraphics[width=\textwidth]{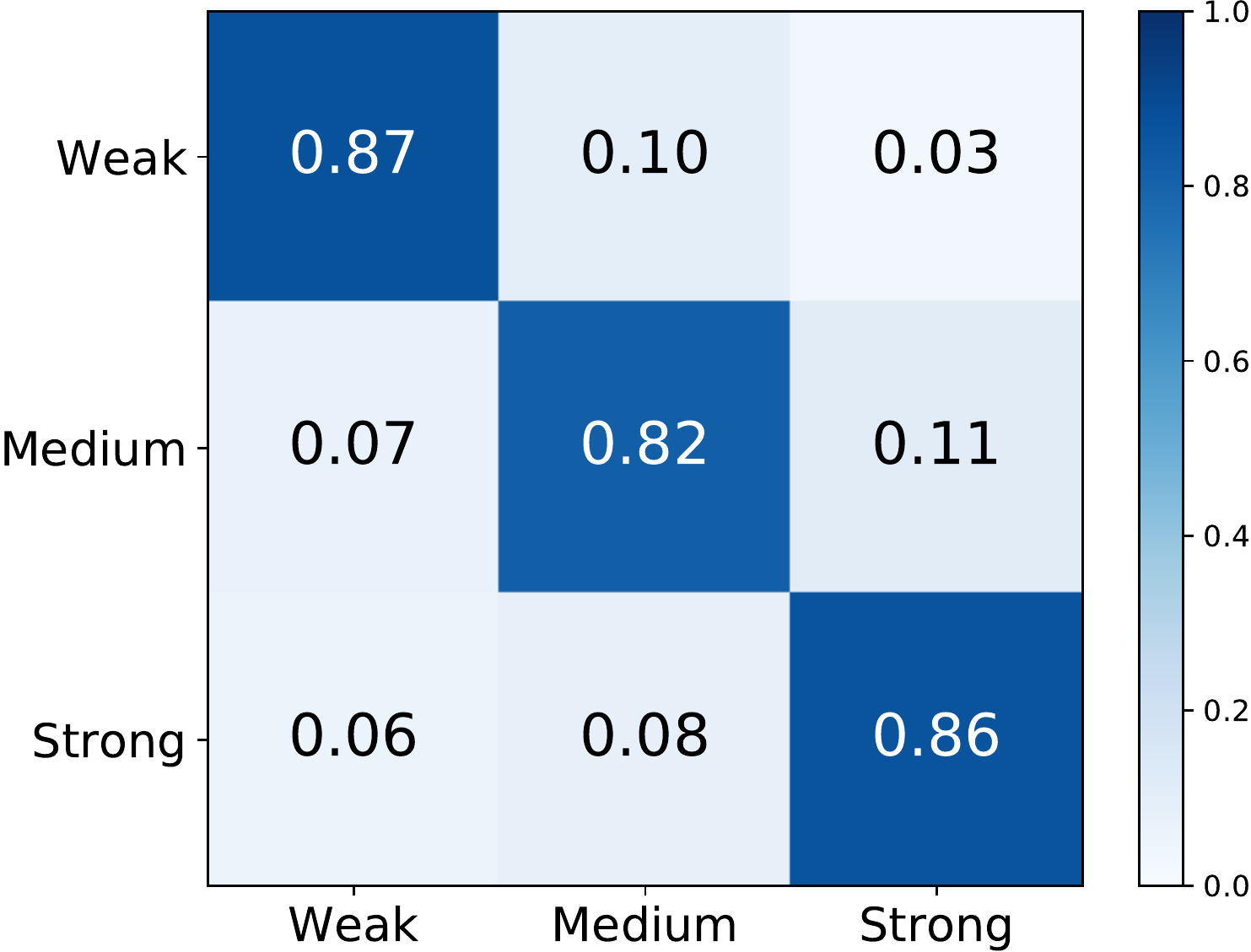}}
			%\centerline{Fear}
		\end{minipage}
        %\centerline{Proposed model}
		\vfill
	\end{minipage}
\end{figure*}

\begin{figure*}[t]
\setlength{\abovecaptionskip}{5pt}
\setlength{\belowcaptionskip}{4pt}
	\centering
    \vspace{-0.4cm}
	\begin{minipage}{\linewidth}
		\begin{minipage}{0.0157\linewidth}
			\centerline{\includegraphics[width=\textwidth]{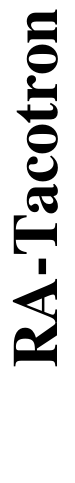}}
		\end{minipage}
		\hfill			
		\begin{minipage}{0.158\linewidth}
			\centerline{\includegraphics[width=\textwidth]{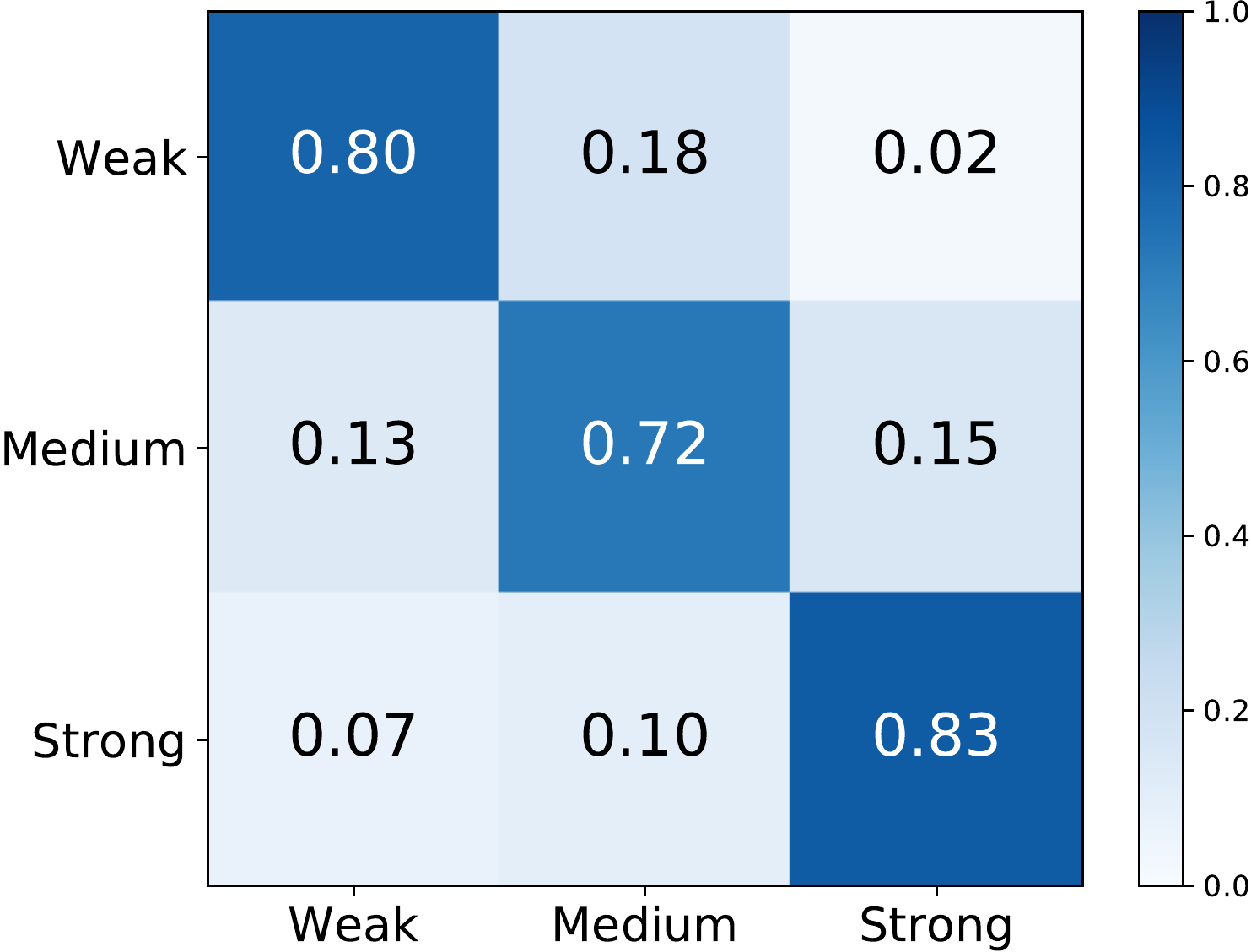}}
			\centerline{Fear}
		\end{minipage}
		\hfill
		\begin{minipage}{0.158\linewidth}
			\centerline{\includegraphics[width=\textwidth]{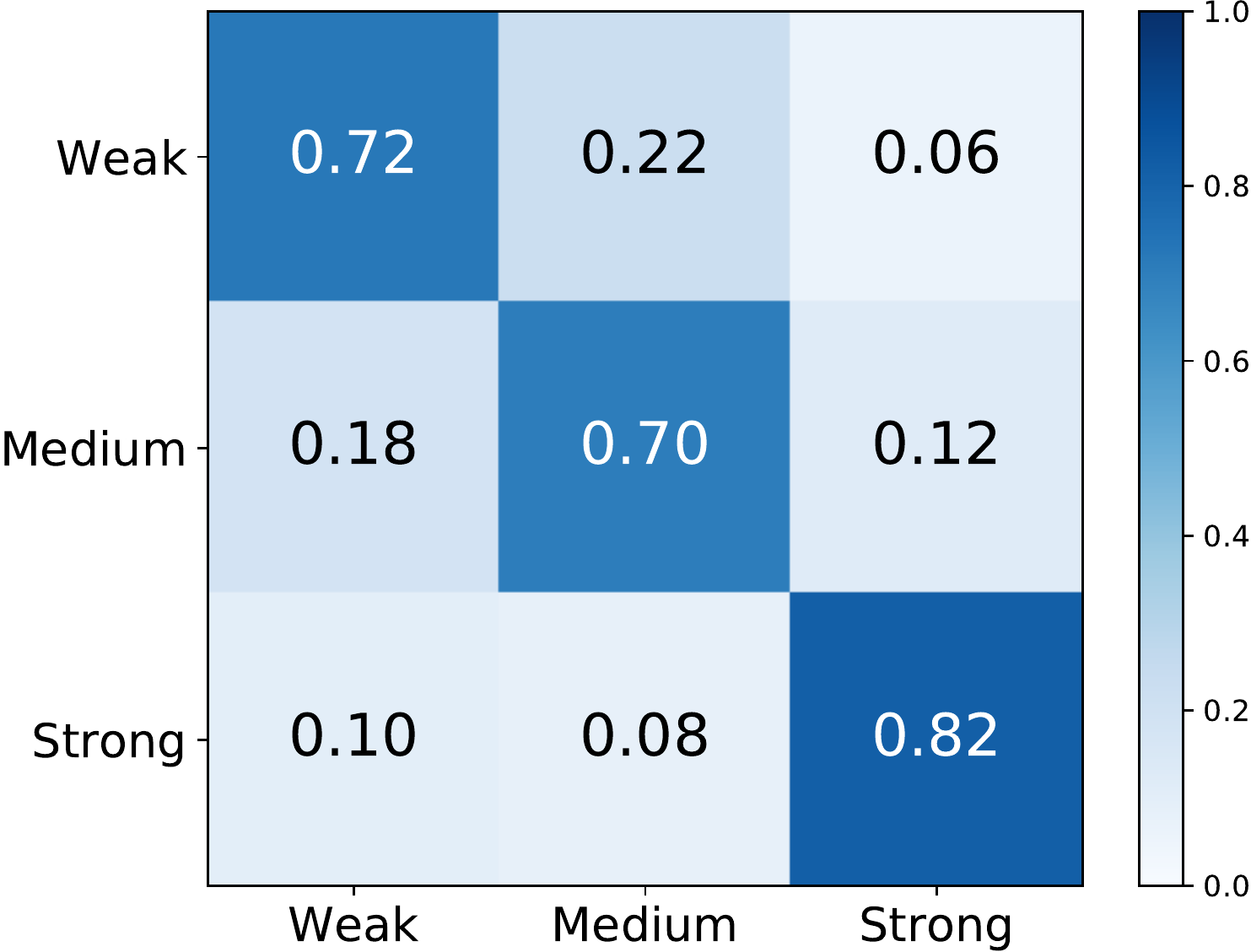}}
			\centerline{Disgust}
		\end{minipage}
		\hfill
		\begin{minipage}{0.158\linewidth}
			\centerline{\includegraphics[width=\textwidth]{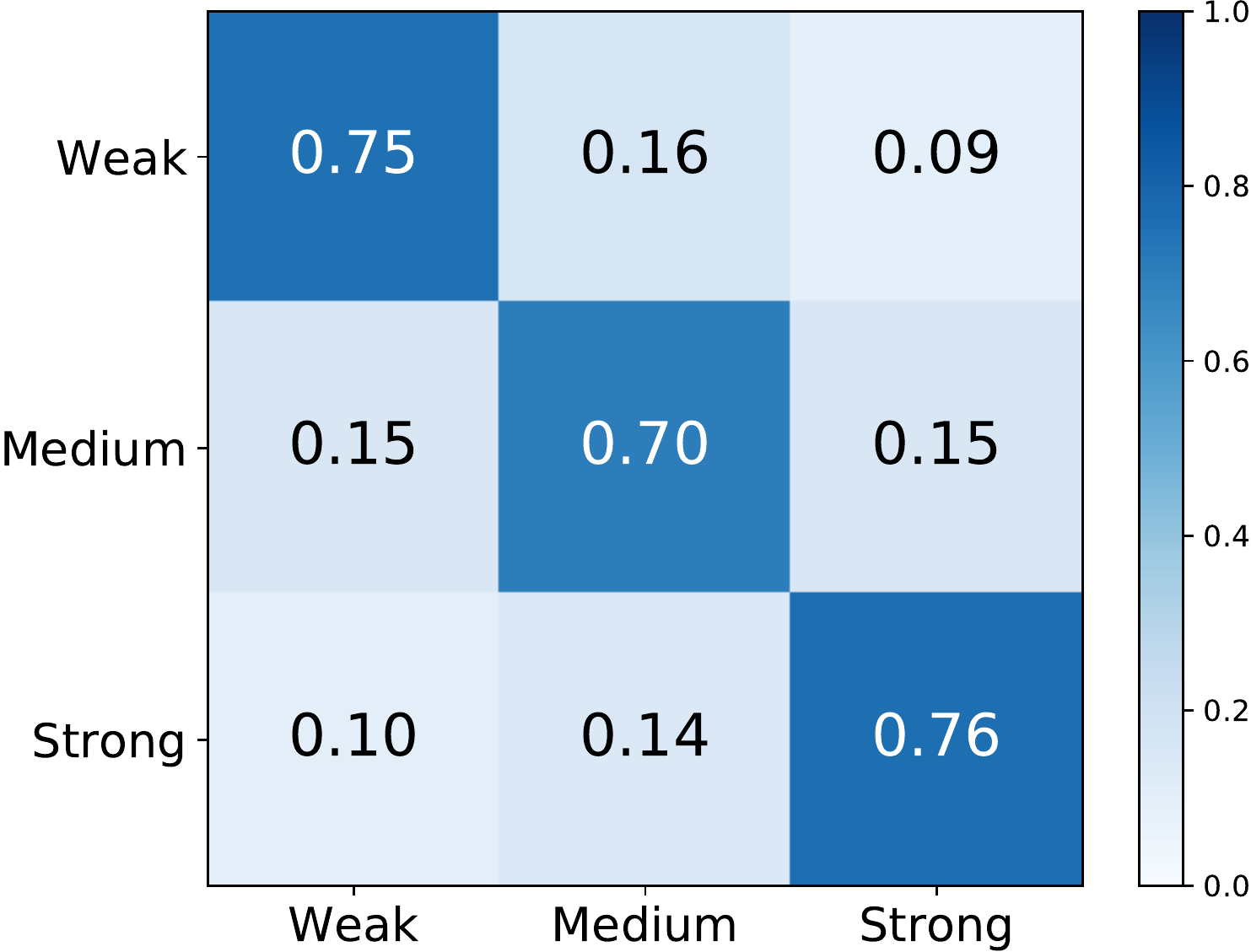}}
			\centerline{Angry}
		\end{minipage}
		\hfill
		\begin{minipage}{0.158\linewidth}
			\centerline{\includegraphics[width=\textwidth]{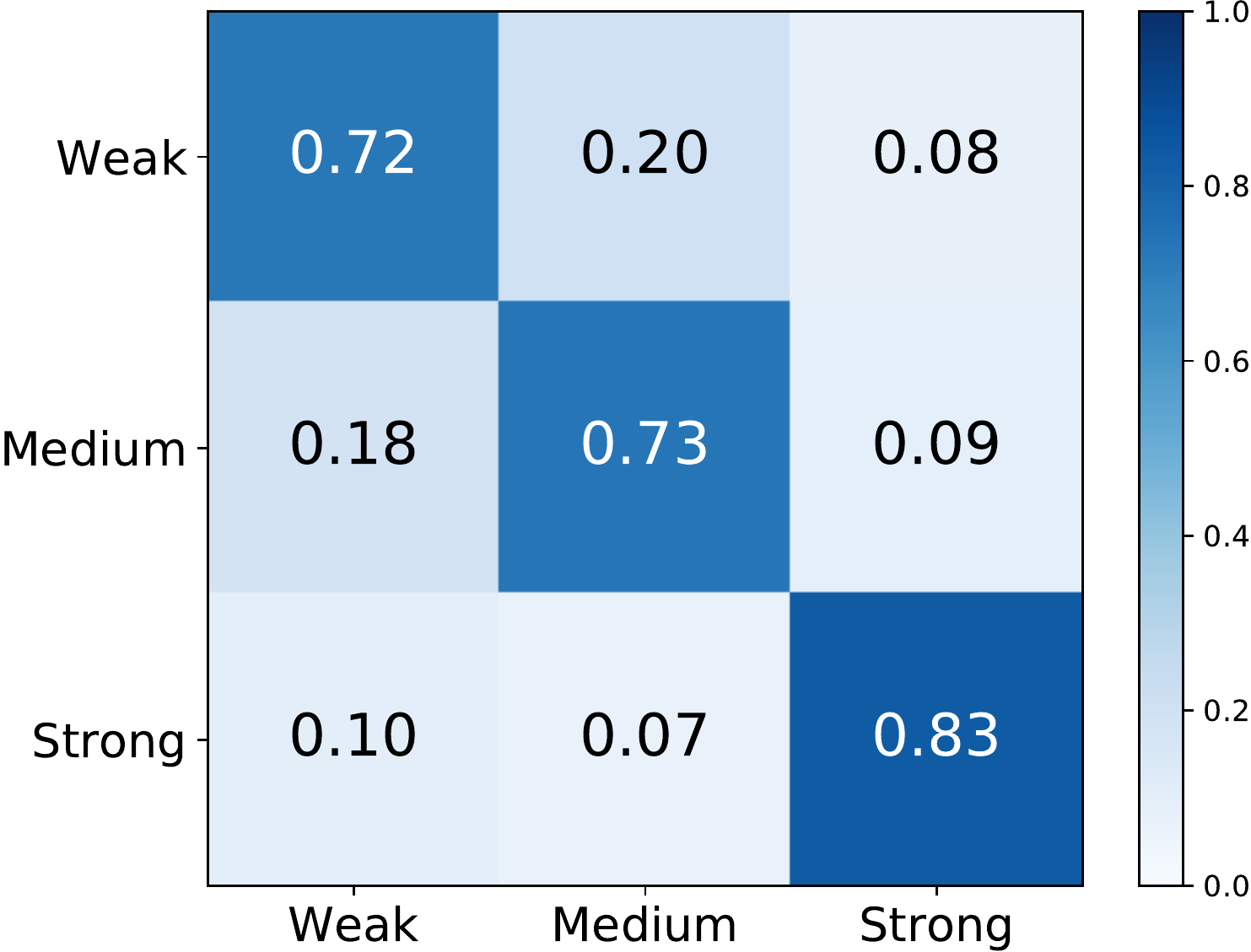}}
			\centerline{Sadness}
		\end{minipage}
		\hfill
		\begin{minipage}{0.158\linewidth}
			\centerline{\includegraphics[width=\textwidth]{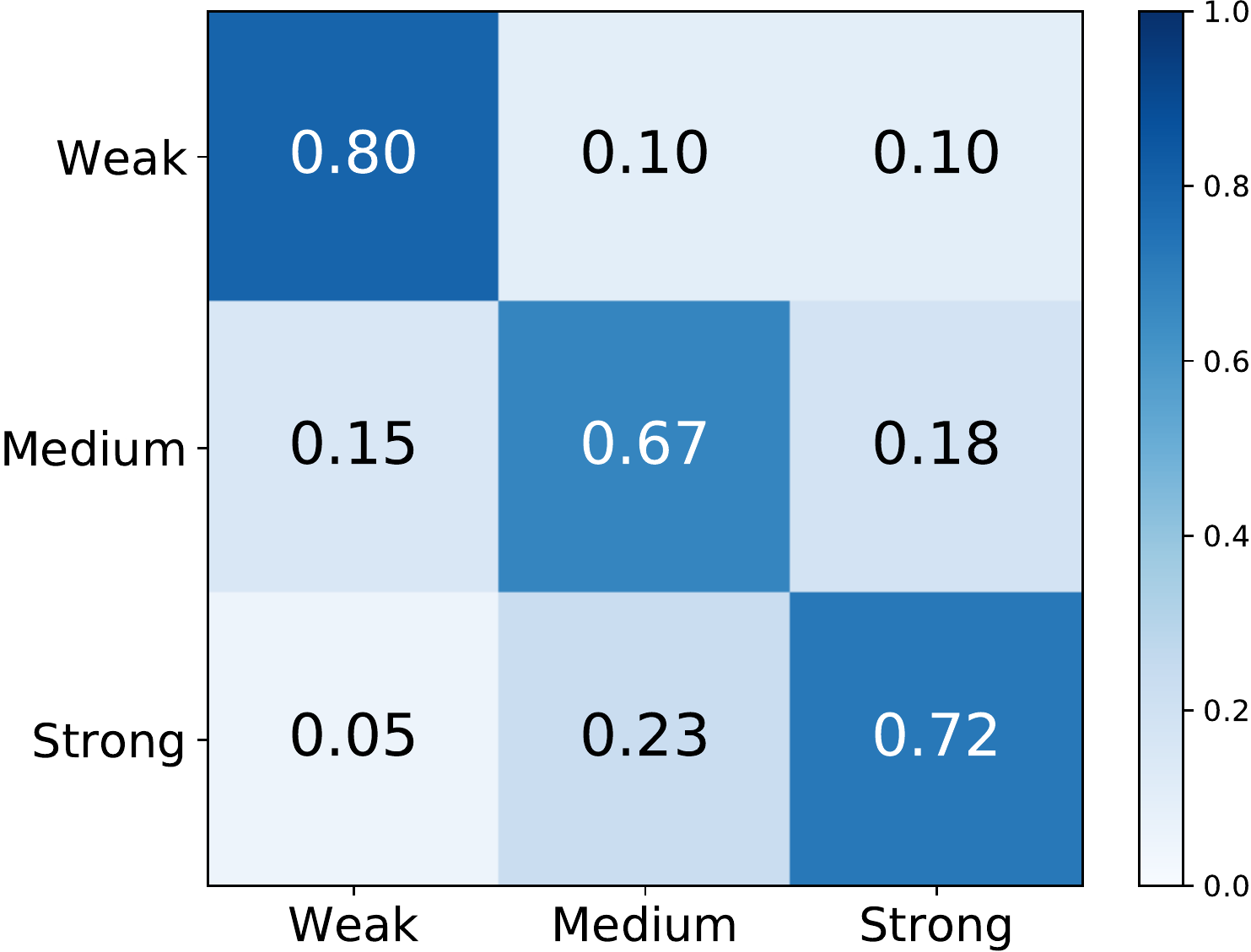}}
			\centerline{Happiness}
		\end{minipage}
		\hfill
		\begin{minipage}{0.158\linewidth}
			\centerline{\includegraphics[width=\textwidth]{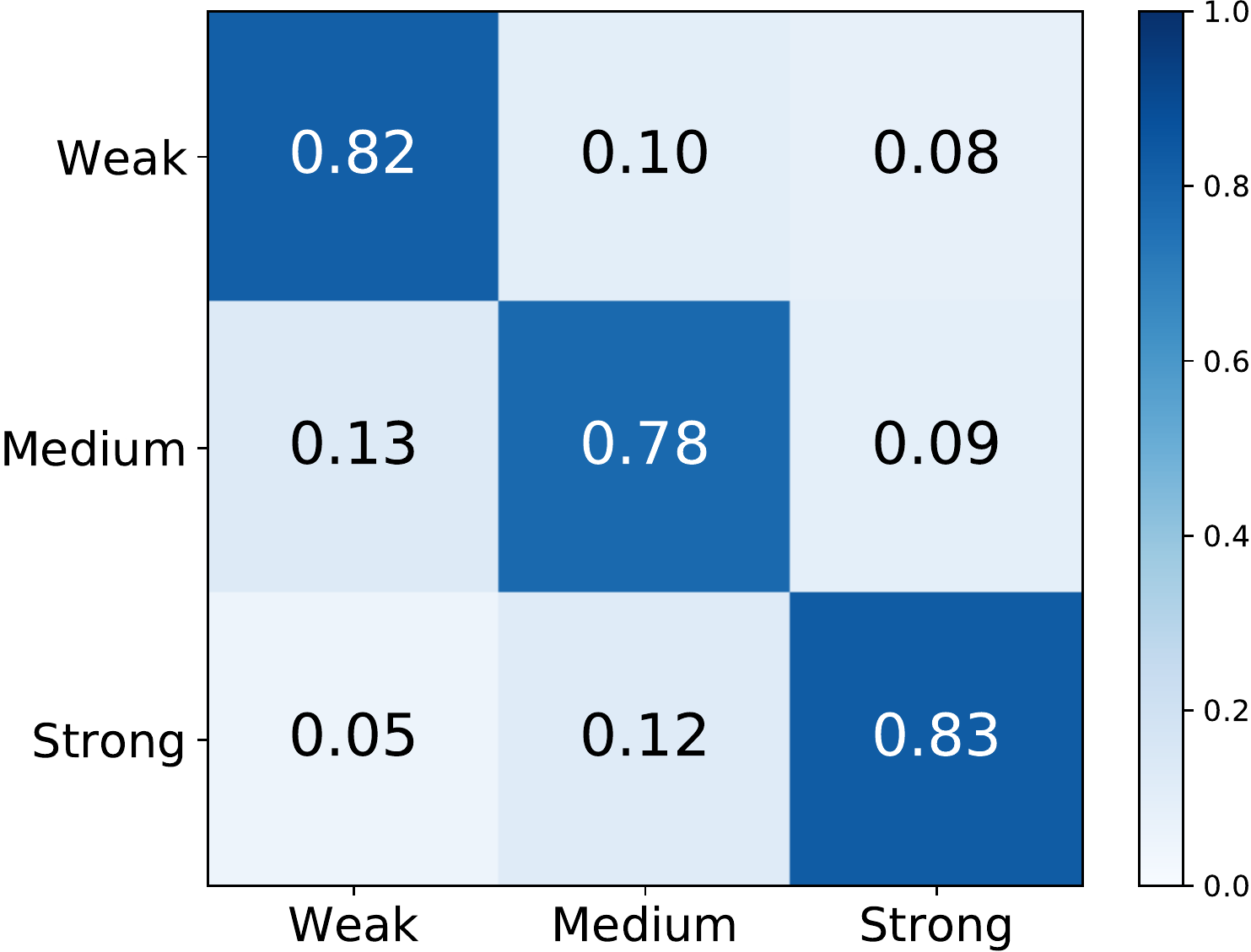}}
			\centerline{Surprise}
		\end{minipage}
        %\centerline{RA-Tacotron2}
		\vfill
	\end{minipage}
    \captionsetup{belowskip=-10pt}
    \caption{Confusion matrices of synthesized speech from the \textbf{Proposed model} (\textbf{the top row}) and the \textbf{RA-Tacotron}~\cite{Zhu2019ControllingES} (\textbf{the bottom row}). The X-axis and Y-axis of subfigures represent perceived and truth emotion strength, respectively.}
    \vspace{-10pt}
    \label{fig:confuse}
\end{figure*}

Style loss~\cite{Johnson2016Perceptual} was first proposed in computer vision to capture the artistic style of an image using the Gram matrix of features maps generated by a CNN, where the Gram matrix computes patch-level appearance statistics, such as texture, in a location-invariant manner. Recently, the Gram matrix has been adopted to measure mel-spectrogram of audio signals~\cite{Ma2019NeuralTS}, aiming to capture local statistics of an audio signal in the frequency-time domain. It is believed that the Gram matrix is able to represent low-level characteristics of speech, e.g. loudness, stress, speed, pitch, etc, which are highly related to emotion expressions.

%There is a lot of work that studies prosody based on spectral characteristics ~\cite{Dmitry Ulyanov (2016); Barry & Kim (2018)}, among which emotions are captured by local statistics in the time-frequency domain~\cite{nature tts Cheang and Pell (2008)}. So we use the method of image transfer to capture and quantify its emotion information from mel spectrogram. Johnson et al.~\cite{Johnson2016Perceptual} use Gram matrix~\cite{GatysA}  of features maps produced by a CNN to capture the style of reference image. The Gram matrix is the inner product of feature map, which represents the correlation between style features.

The emotion embedding in our proposed model is a collection of CNN output sequences, which can be naturally seen as the feature map of mel-spectrogram. Each value of feature map from the convolution of a specific filter at a target location, and the essence is the extraction and quantification of features, so each value can be seem as the strength of emotion-related features. Our goal is to synthesize speech with a certain target emotion category (such as surprise), while flexibly controlling the strength of the emotion transferred to the target. To achieve this, the emotion \textit{style} difference between the generated speech and the reference speech can be measured using style loss. Specifically, given the emotion embedding (feature map) of reference and synthesized speech $R$ and $S$, their corresponding gram matrices $G$ and $I$ are calculated by inner-product as:

\begin{equation}
\setlength{\abovedisplayskip}{3pt}
\setlength{\belowdisplayskip}{3pt}
G=R^T*R,  \ \ \ \text{and} \quad I=S^T*S,
\label{eq1}
\end{equation}
That is to say, style information is measured as the amount of correlation present between features maps. The Gram matrix essentially captures the distribution of features of a set of feature maps. By trying to minimize the style loss between two gram matrices, we are essentially matching the distribution of features between the two emotion embeddings. Specifically, the style loss $L_{sty}$ minimizes the $L2$ distance between two Gram matrices, making the synthesized speech as close as possible to the reference audio in style of emotion:

\begin{equation}
\setlength{\abovedisplayskip}{3pt}
\setlength{\belowdisplayskip}{3pt}
L_{sty} = \frac1{(2NM)^2}\sum_{}(I-G)^2,
\label{eq2}
\end{equation}
where $N$ and $M$ are the number of rows and columns of the matrix respectively. Finally, the total loss of the proposed model becomes:
\begin{equation}
\setlength{\abovedisplayskip}{3pt}
\setlength{\belowdisplayskip}{3pt}
\begin{split}
%\label{E1}
L_{total}=L_{tac}+L_{sty}+L_{cls\_src}+L_{cls\_tgt},
\end{split}
\end{equation}
where $L_{tac}$ is the typical Tacotron MSE loss, $L_{sty}$ is the style loss in Eq.~(\ref{eq2}), and $L_{cls\_src}$ and $L_{cls\_tgt}$ are classification loss of the emotion embedding network and the auxiliary classifier network, respectively.

\vspace{-5pt}
\subsection{Emotion strength control}
\vspace{-2pt}
Since the emotion embedding can be viewed as the feature map of the mel spectrogram, representing the strength of emotion-related features, the emotion strength in the synthetic speech can be controlled easily by adjusting the value of emotion embedding. In this work, we use an emotion scalar to multiply the emotion embedding to control emotion transfer strength at the inference stage, as shown in Figure~\ref{fig:frame_diagram}. This is similar to the degree control in image style transfer~\cite{Johnson2016Perceptual}.  Note that this scalar always equals to 1 during the training process.

\vspace{-5pt}
\section{Experiments}
\label{sec:typestyle}
We evaluate the performance of our proposed model in emotion transfer and emotion strength control through subjective evaluations and sample analysis. Twenty (gender balanced) native Mandarin listeners are invited to participate in the evaluation.

\vspace{-5pt}
\subsection{Experimental setup}
\vspace{-2pt}
We use the same dataset as in~\cite{Zhu2019ControllingES}: a high-quality emotional speech corpus containing 14-hour of recordings by a professional Chinese actress. She imitates a little girl to speak with seven categories of emotion ($neutral$, $happy$, $angry$, $disgust$, $fear$, $surprise$ and $sadness$). There are 6000 sentences in the $neutral$ emotion category, and 620 sentences in each of the remaining emotion categories. During model training, all the recordings are down-sampled from 44 kHz to 16 kHz. We randomly select 10 sentences from each emotion data as the subjective listening test set.

During inference, we randomly select one sentence from each emotion test set as the emotion reference audio. We conduct experiments on both emotion transfer and emotion strength control. For the latter experiments, although the emotion scalar can be controlled continuously to represent emotion strength in each emotion category, we set it to 0.5, 1.5 and 2.5 to represent three typical emotion strengths -- weak, medium and strong for subjective comparison. Even though the larger scalar means the stronger emotion strength, the scalar cannot be infinite. We find that when the strength scalar is greater than 3, it will lead to the excessive transfer of emotion. For example, the generated speech with anger emotion will have a very fast speaking rate which affects intelligibility. In particular, when the emotion scalar is as low as 0.1, the target emotion will change to neutral speech. As a result, in the emotion strength control experiments, each test sentence will be synthesized to 18 samples, including 6 kinds of emotions (without $neutral$) and each has three emotion strengths.

\begin{figure*}[t]
    \setlength{\abovecaptionskip}{4pt}
    \setlength{\belowcaptionskip}{3pt}
	\centering
	\centerline{\includegraphics[width=0.48\textwidth]{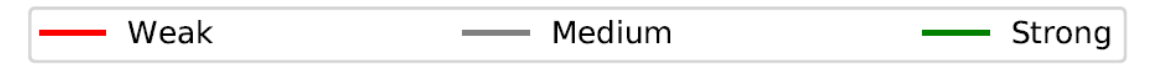}}
	\begin{minipage}{\linewidth}
		
		\begin{minipage}{0.162\linewidth}
			\centerline{\includegraphics[width=\textwidth]{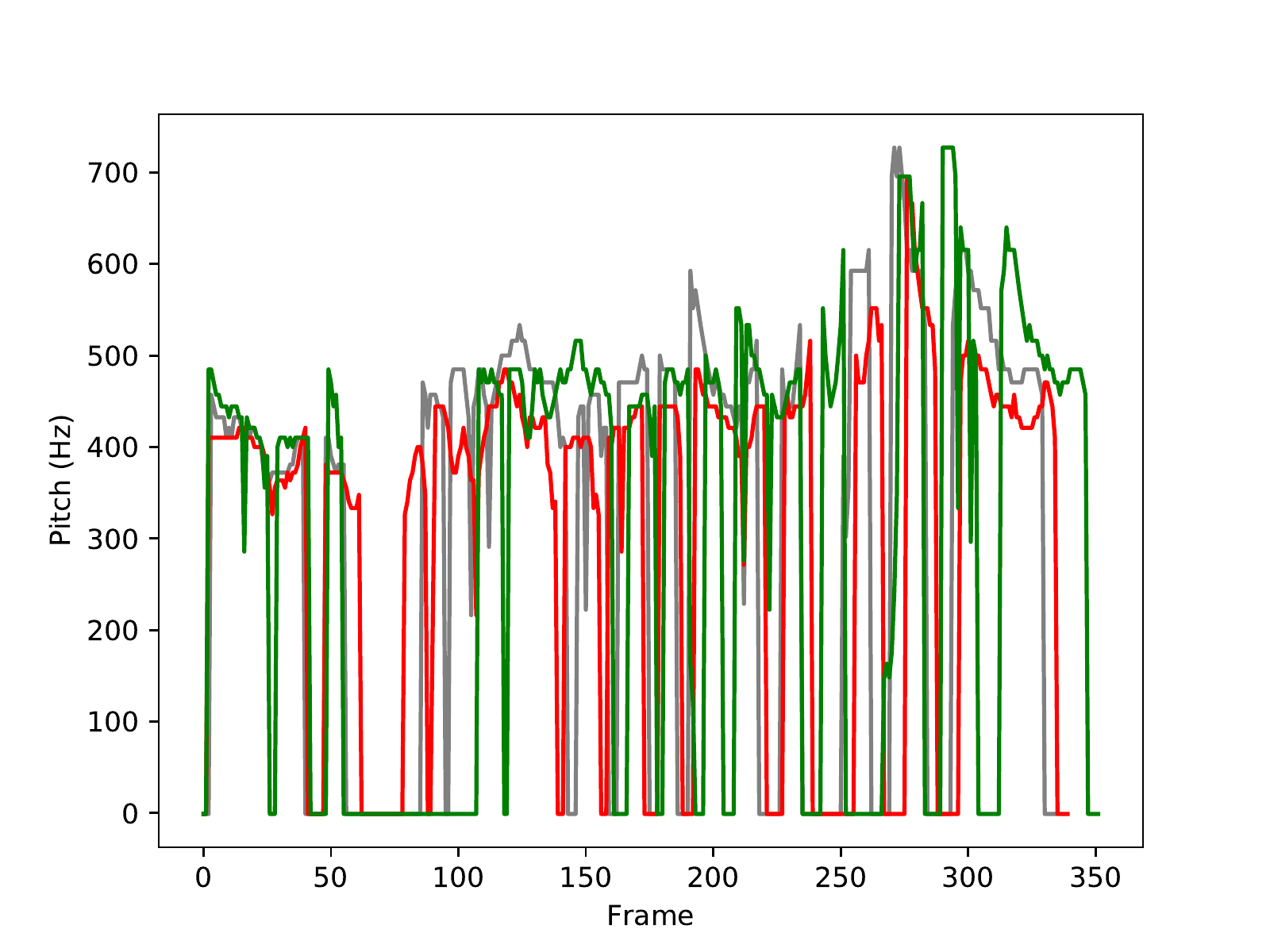}}
			\centerline{Fear}
		\end{minipage}
		\hfill
		\begin{minipage}{0.162\linewidth}
			\centerline{\includegraphics[width=\textwidth]{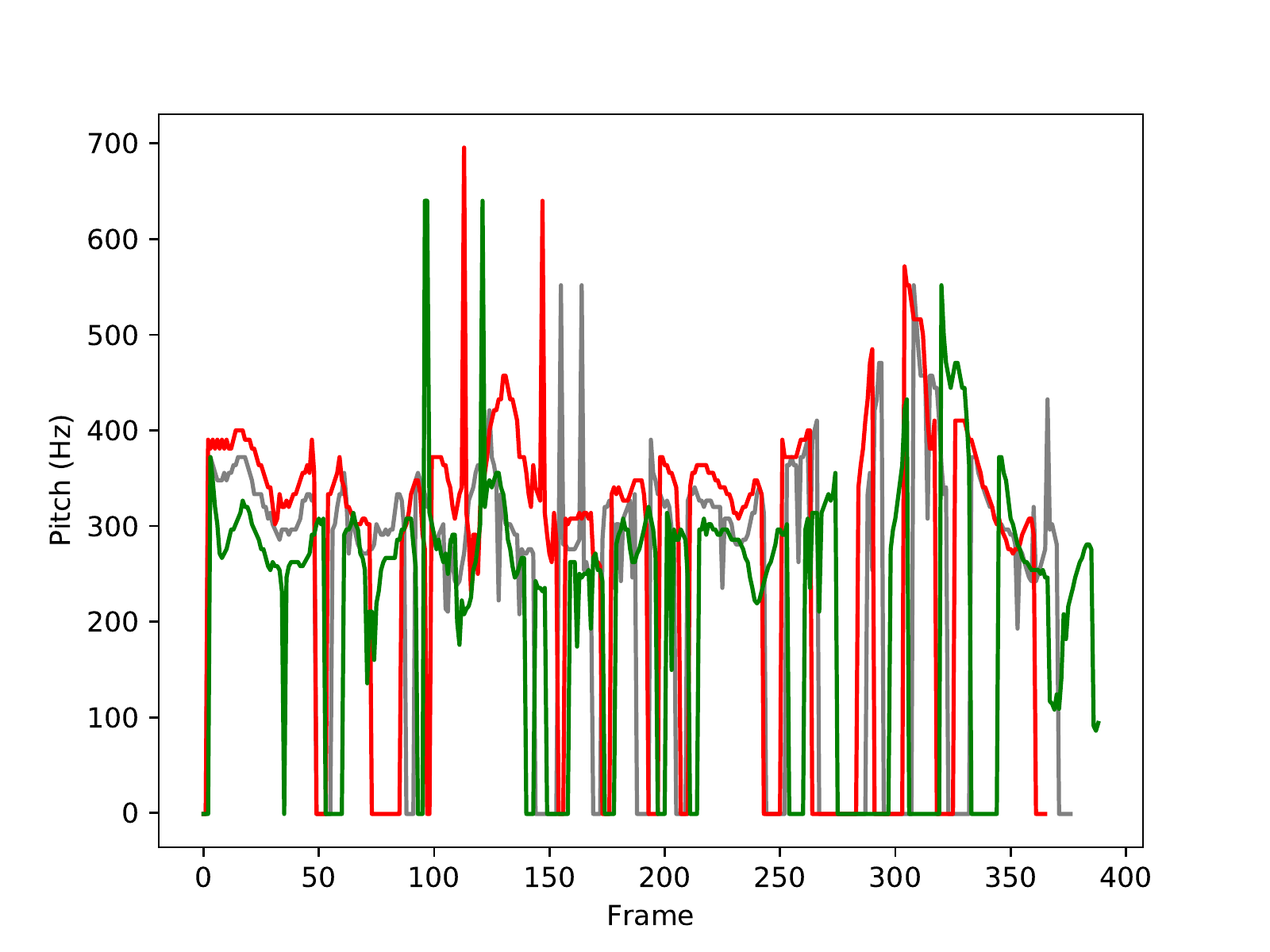}}
			\centerline{Disgust}
		\end{minipage}
		\hfill
		\begin{minipage}{0.162\linewidth}
			\centerline{\includegraphics[width=\textwidth]{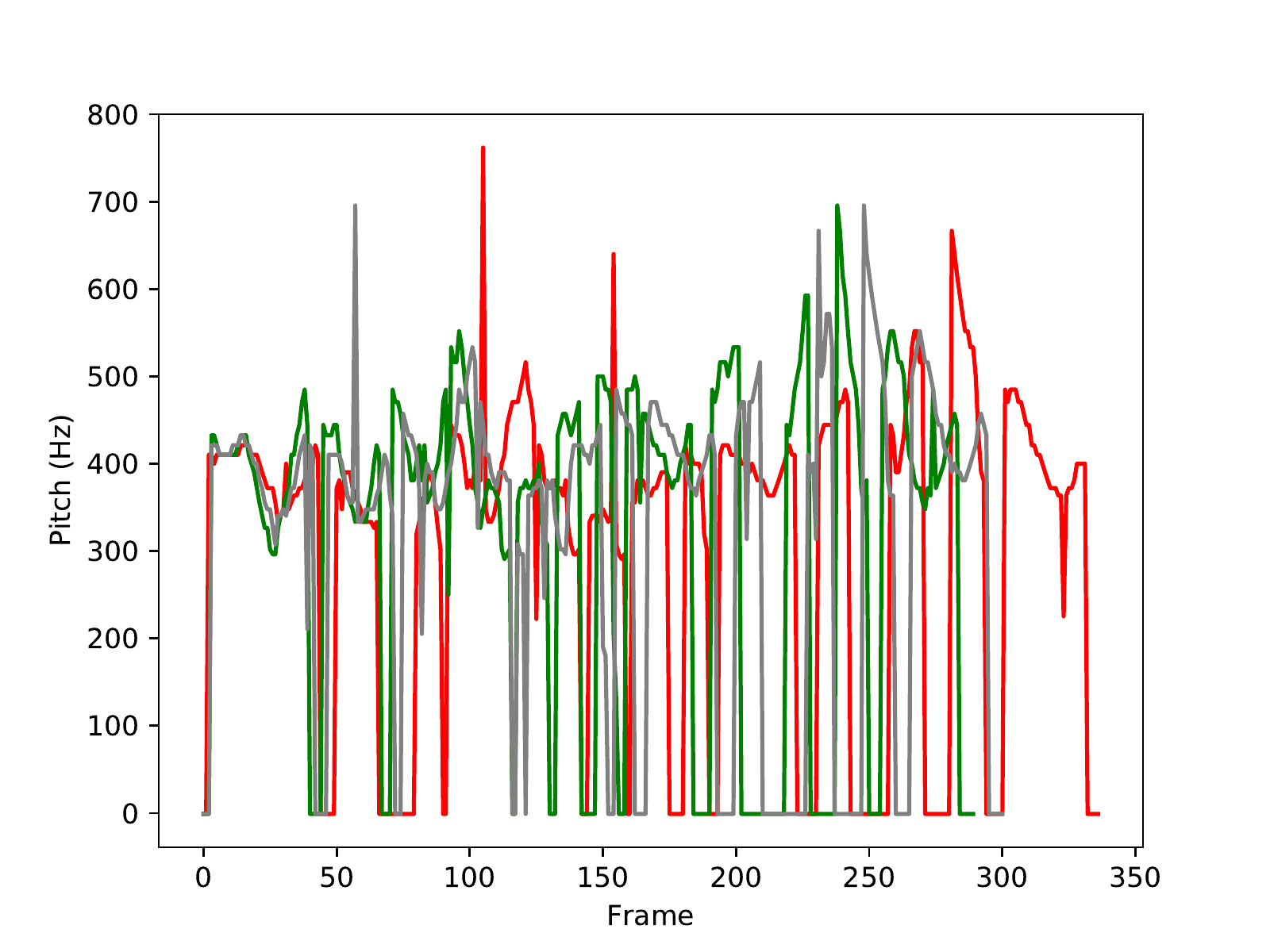}}
			\centerline{Angry}
		\end{minipage}
		\hfill
		\begin{minipage}{0.162\linewidth}
			\centerline{\includegraphics[width=\textwidth]{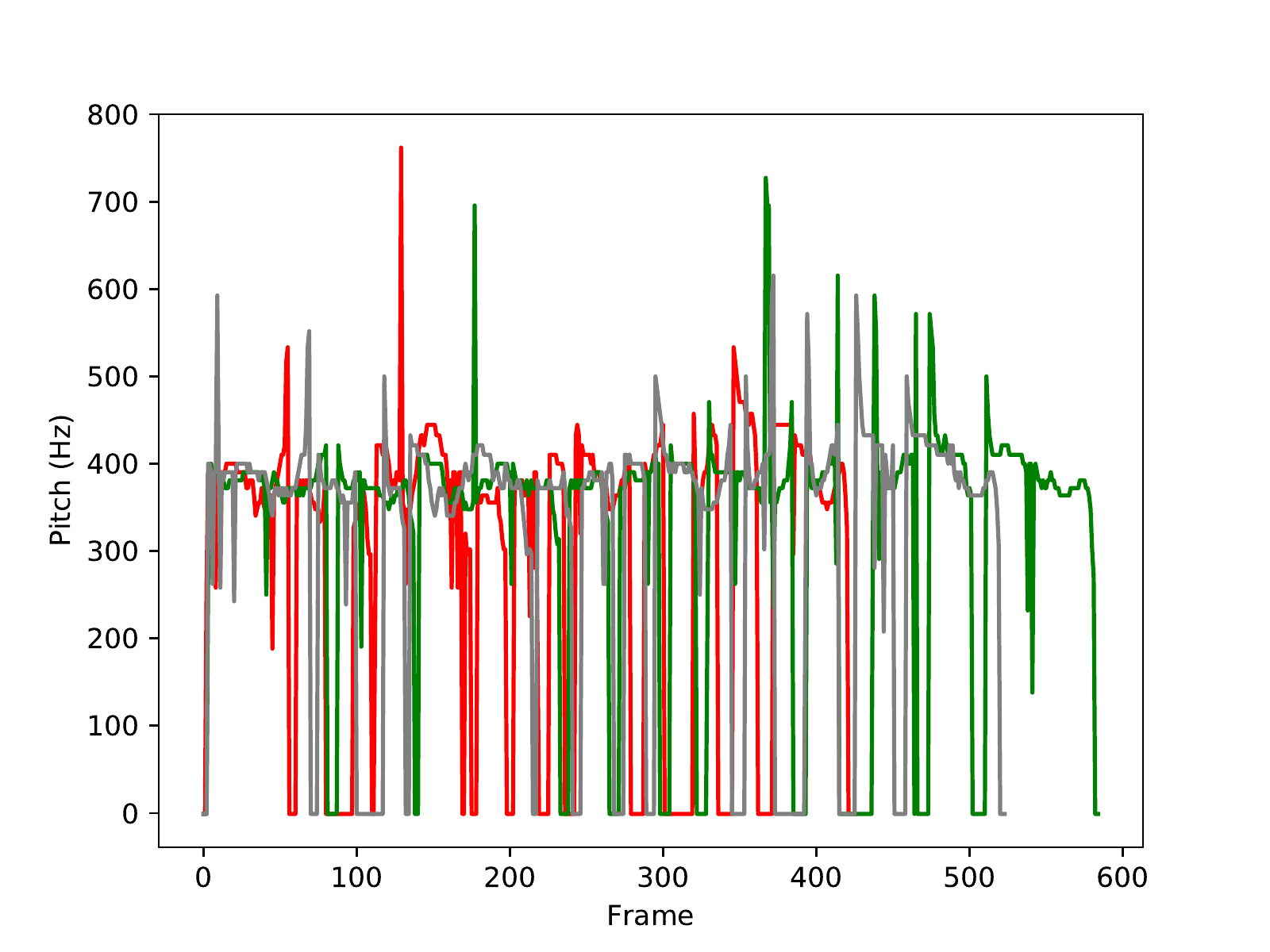}}
			\centerline{Sadness}
		\end{minipage}
		\hfill
		\begin{minipage}{0.162\linewidth}
			\centerline{\includegraphics[width=\textwidth]{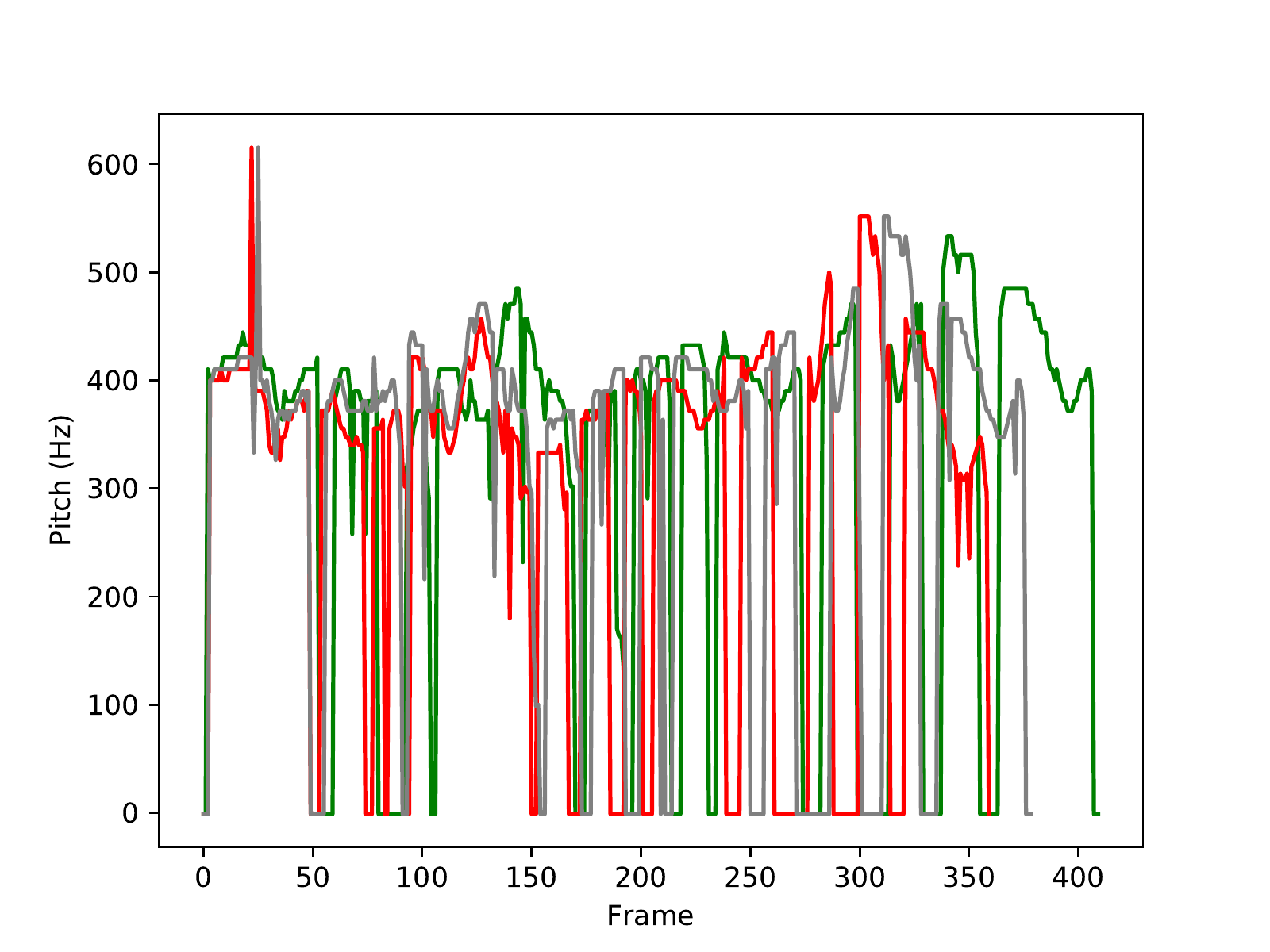}}
			\centerline{Happy}
		\end{minipage}
		\hfill
		\begin{minipage}{0.162\linewidth}
			\centerline{\includegraphics[width=\textwidth]{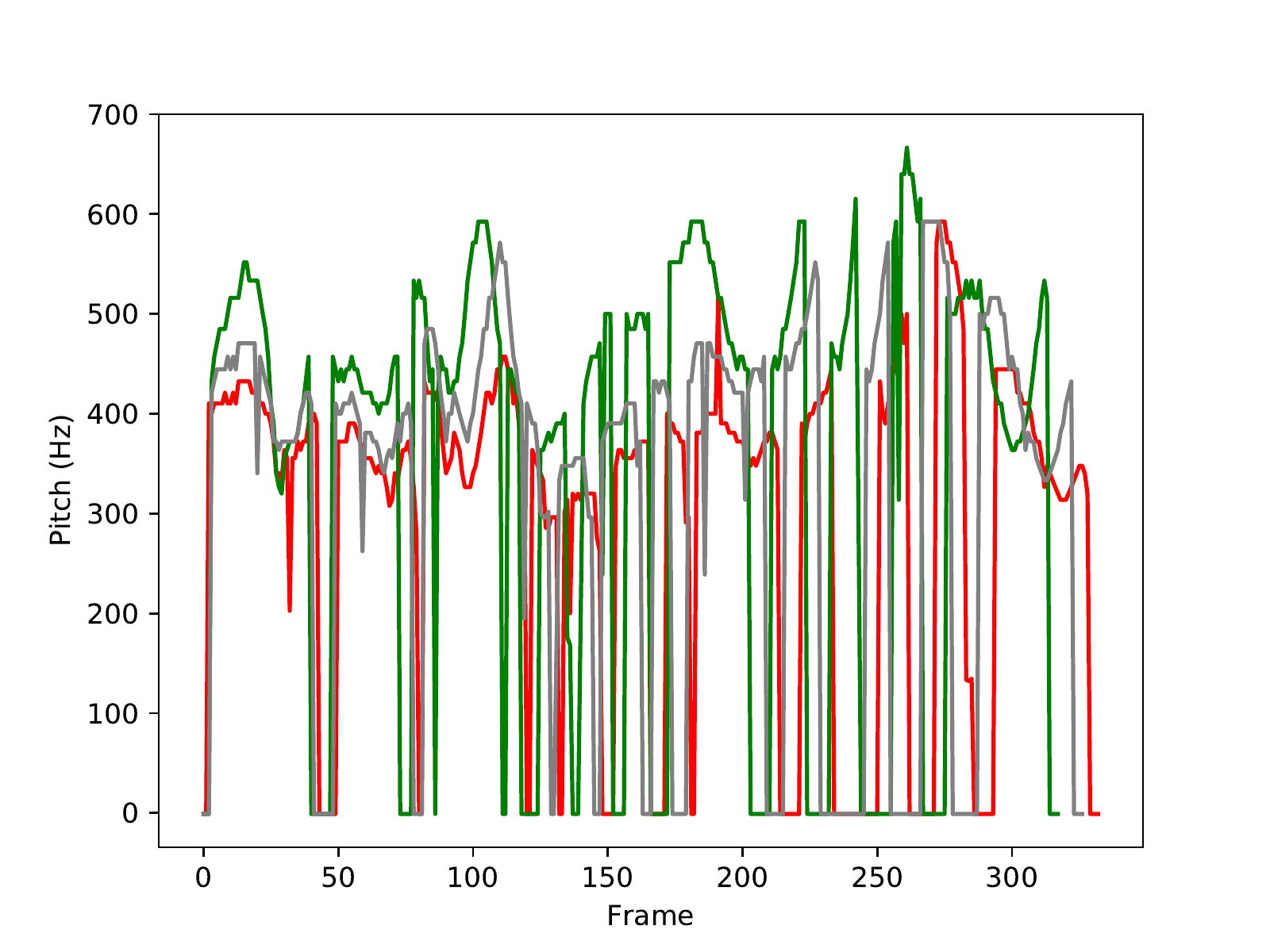}}
			\centerline{Surprise}
		\end{minipage}
		\vfill
	\end{minipage}
    \captionsetup{belowskip=-10pt}
    \caption{Pitch contours of synthesized speeches with the same text for six emotions and three emotion strength.}
    \label{fig:pitch}
\end{figure*}

\begin{figure*}[t]
\setlength{\abovecaptionskip}{3pt}
\setlength{\belowcaptionskip}{-5pt}
	\centering
	\begin{minipage}{\linewidth}
		
		\begin{minipage}{0.162\linewidth}
			\centerline{\includegraphics[width=\textwidth]{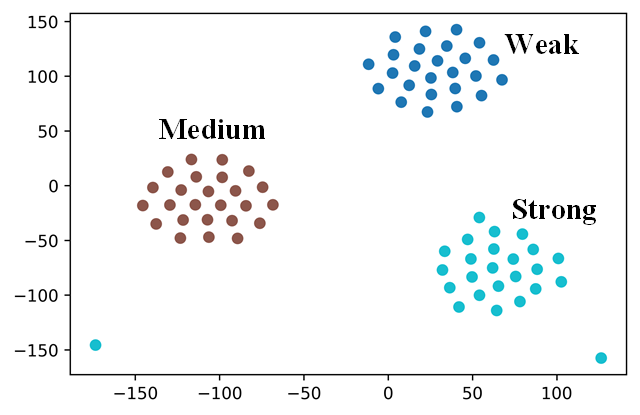}}
			\centerline{Fear}
		\end{minipage}
		\hfill
		\begin{minipage}{0.162\linewidth}
			\centerline{\includegraphics[width=\textwidth]{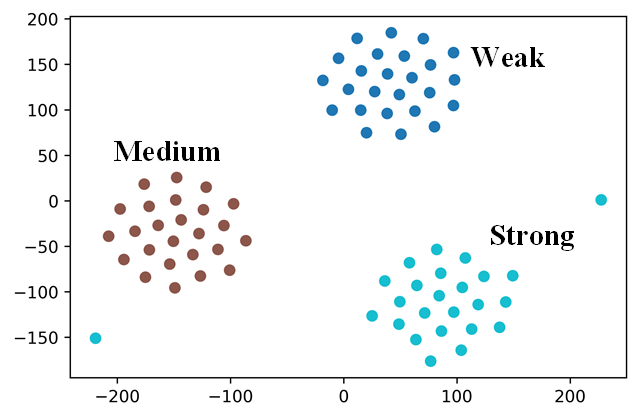}}
			\centerline{Disgust}
		\end{minipage}
		\hfill
		\begin{minipage}{0.162\linewidth}
			\centerline{\includegraphics[width=\textwidth]{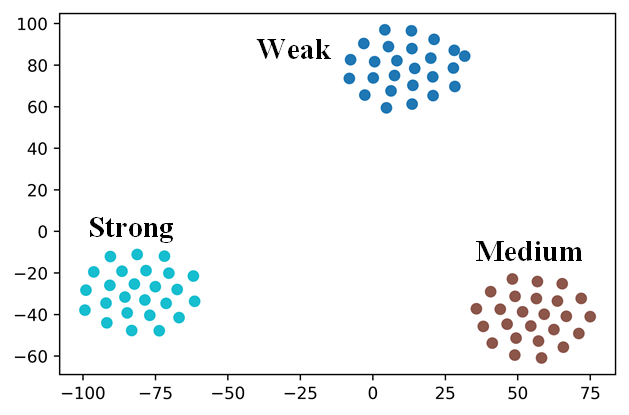}}
			\centerline{Anger}
		\end{minipage}
		\hfill
		\begin{minipage}{0.162\linewidth}
			\centerline{\includegraphics[width=\textwidth]{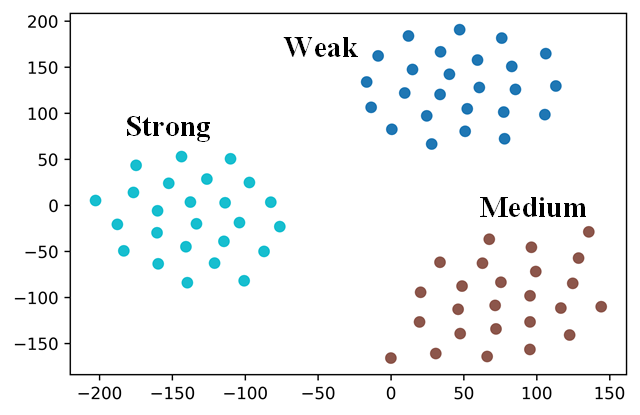}}
			\centerline{Sadness}
		\end{minipage}
		\hfill
		\begin{minipage}{0.162\linewidth}
			\centerline{\includegraphics[width=\textwidth]{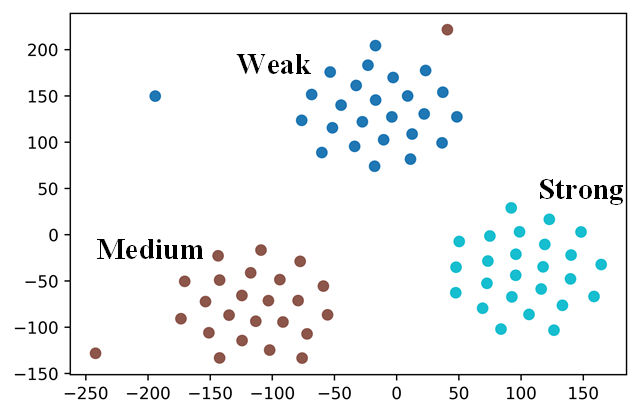}}
			\centerline{Happy}
		\end{minipage}
		\hfill
		\begin{minipage}{0.162\linewidth}
			\centerline{\includegraphics[width=\textwidth]{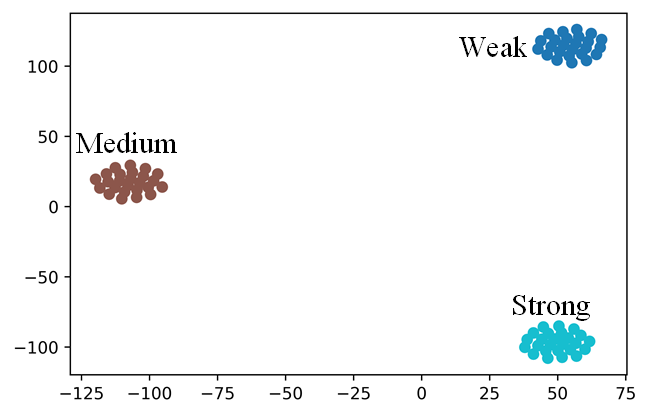}}
			\centerline{Surprise}
		\end{minipage}
		\vfill
	\end{minipage}
    %\captionsetup{belowskip=-15pt}
    \caption{t-SNE visualization of three emotion strengths for six emotion categories. The values of emotion strength scalar for three emotion strengths are : 0.5 (weak), 1.5 (medium), 2.5 (strong).}
    \vspace{-14pt}
    \label{fig:strength tsne}
\end{figure*}

\vspace{-7pt}
\subsection{Emotion transfer}
\vspace{-2pt}
\subsubsection{Ablation studies}
\vspace{-3pt}

We first conduct ablation studies on emotion transfer to validate the effectiveness of different structures with different losses. Specifically, we conduct a subjective emotion classification test on 10 synthetic samples for each emotion where the sentences are chosen form the $neutral$ test set in order to let the listeners to focus on the emotion expression delivered by audio instead of the text. The emotion scalar is set to 1 as we do not evaluate the strength control in this experiment.

\begin{table}[H]
  \vspace{-8pt}
  \setlength{\abovecaptionskip}{2pt}
  \setlength{\belowcaptionskip}{3pt}
  %\caption{Subjective emotion category classification results of ablation studies on our proposed model}
  \caption{Accuracy of subjective emotion category classification based on the ablation study of our proposed model.}
  \label{tab:mos}
  \centering
  \renewcommand\arraystretch{1}
  \begin{tabular}{p{1cm}|p{0.5cm}<{\centering}<{\centering}|p{1cm}<{\centering}|p{1cm}<{\centering}|p{1cm}<{\centering}|p{0.5cm}<{\centering}}
  \toprule %\hline
  Loss & \textbf{$L_{tac}$} & \textbf{+$L_{cls\_tgt}$} & \textbf{+$L_{cls\_src}$} & \textbf{+$L_{cls\_src}$ +$L_{cls\_tgt}$} & \textbf{$L_{total}$} \\ \hline % & \textbf{$L_{total}$}
    fear      &0.71  & 0.76  &0.91  & 0.95  & 0.97 \\
    disgust   &0.67  & 0.73  &0.75  & 0.79  & 0.85 \\
    angry     &0.81  & 0.94  &0.92  & 0.96  & 0.98 \\
    sadness   &0.95  & 0.93  &0.96  & 0.98  & 0.97 \\
    happy     &0.66  & 0.70  &0.72  & 0.75  & 0.84 \\
    surprise   &0.62  & 0.67  &0.75  & 0.78  & 0.82 \\
  \bottomrule
  \end{tabular}
  \vspace{-6pt}
  %\captionsetup{belowskip=-40pt}
  \end{table}

Classification accuracy is shown in Table~\ref {tab:mos}. Note that there are 200 listening samples for each emotion category and listeners are asked to select one from the 6 emotion categories for each testing sample. We can see from Table~\ref {tab:mos} that the $sad$ emotion is the easiest to distinguish as its samples always have very low pitch, energy and slow speaking speed. We also notice that the addition of the emotion classification losses ( $L_{cls\_tgt}$ and $L_{cls\_src}$) and the style loss ($L_{sty}$) can bring substantial gain for the classification accuracy. The combination of all losses achieves the best accuracy, which outperforms the baseline Prosody-Tacotron~\cite{Skerry2018Towards} (only $L_{tac}$) by a large margin.

\begin{figure}[H]
\setlength{\abovecaptionskip}{-10pt}
\setlength{\belowcaptionskip}{-9pt}
\begin{minipage}[b]{1.0\linewidth}
  \centering
  \vspace{-0.3cm}
  \centerline{\includegraphics[height=4.5cm,width=7.5cm]{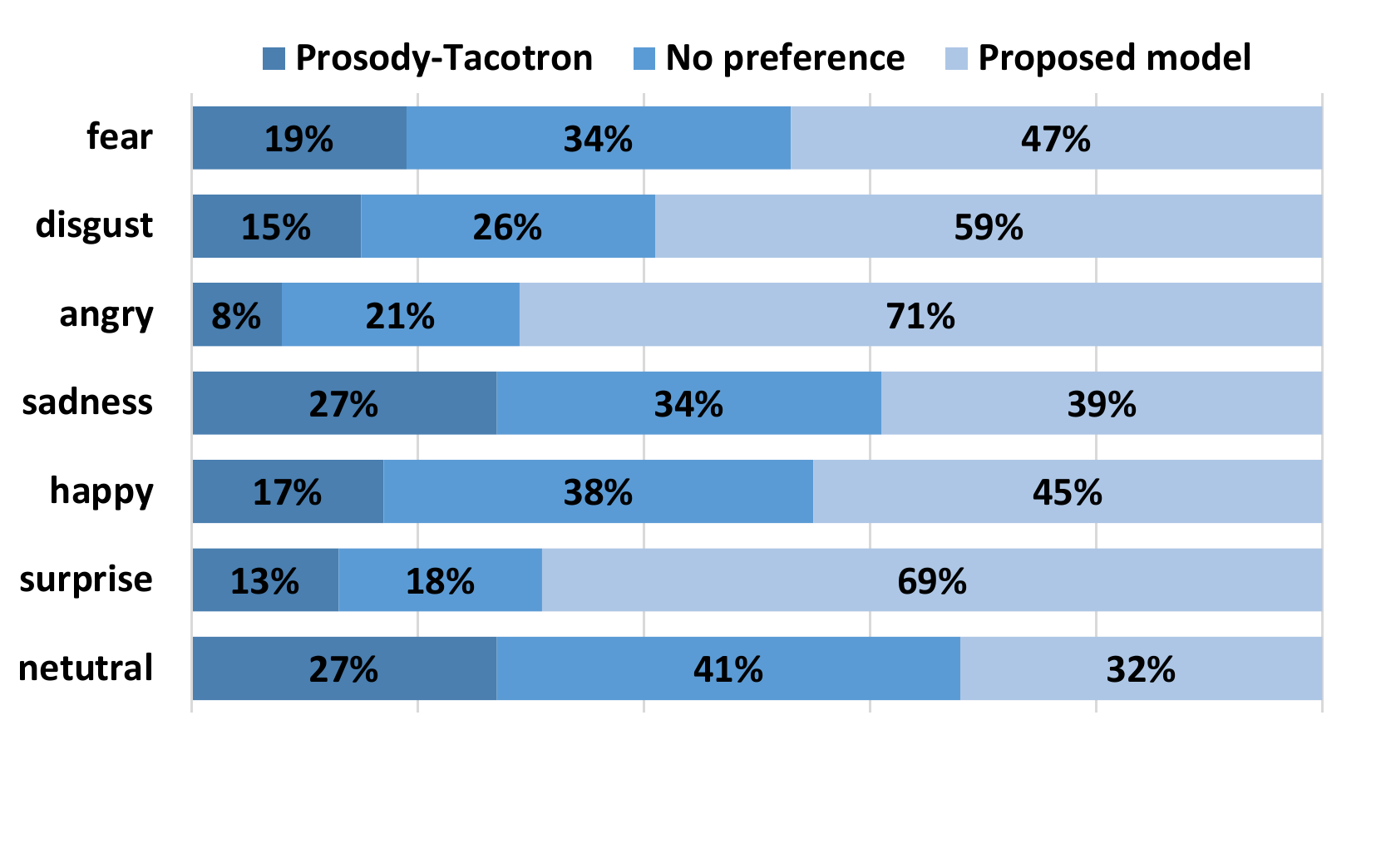}}
 \vspace{-4pt}
%  \centerline{(a) Result 1}\medskip
  \caption{Preference test results of emotion transfer (emotion scale is 1) with seven emotion categories between the prosody-Tacotron model and proposed model.}
 % \caption{Preference of clarity and delicate about emotion expression.}
 \captionsetup{belowskip=-20pt}
  \label{fig:abtest}
  \vspace{-6pt}
 \end{minipage}
\end{figure}

\vspace{-5pt}
\subsubsection{Preference test}
\vspace{-2pt}

We also conduct a preference test to compare the expressiveness of emotion transfer between the proposed model (use of all losses) and the baseline Prosody-Tacotron~\cite{Skerry2018Towards} (use $L_{tac}$ loss). Listeners are offered a pair of randomly selected samples from two models and asked to choose which one is as expressive as the reference audio. If there is no discernible difference between the two samples, they can choose no preference. The preference test results in Figure~\ref{fig:abtest} show that our proposed model always achieves much more preference than Prosody-Tacotron in all 6 emotion classes, and its emotional expressiveness is much closer to the reference audio, resulting in more expressive emotional speech. The preference gap in $neutral$ is slightly small, and it is because neutral emotion is not salient in prosodic variations.

\vspace{-6pt}
\subsection{Emotion strength control}
\vspace{-1pt}
We further evaluate the ability to control the strength of emotion transfer. Here we compare the proposed method with RA-Tacotron~\cite{Zhu2019ControllingES}, in which a similar scalar is used to control emotion strength. We follow the model configuration as in~\cite{Zhu2019ControllingES} to re-implement the model. We conduct strength ordering test, where listeners need to order by strength the randomly placed samples with different strengths for the same testing sentence. The samples can be played as many times as the listeners want. For the RA-Tacotron, we set scalar to 0, 0.5 and 1 to represent weak, medium and strong, same as~\cite{Zhu2019ControllingES}. Note that strength in RA-Tacotron is normalized to 0-1. Fig~\ref{fig:confuse} presents the strength confusion matrices for the proposed model and RA-Tacotron. We can find that our proposed model has a higher classification accuracy of emotion strength, with less strength confusions as compared with RA-Tacotron. This indicates that the proposed model has finer ability to perform strength control based on the emotional reference audio.

We further analyze the influence of different emotion strengths on pitch trajectory as well as speaking rate on the synthetic emotional samples. As shown in Figure~\ref{fig:pitch}, the pitch trajectories of different strengths within the same emotion follow roughly the same trend, but the details vary widely. Moreover, the changes of emotion strength not only reflect on the pitch but also the speaking rate and the utterance level tone. For example, the pitch, speaking rate, and tone variation of $surprise$ and $anger$ increase with the increase of strength from weak to strong. We also visualize three different strengths for each emotion category in emotion embedding space by t-distributed stochastic neighbor embedding (t-SNE) plots~\cite{Laurens2008Visualizing}. Figure~\ref{fig:strength tsne} shows that different strengths have clearly different clusters. The above case analysis demonstrates that the emotion embedding space can effectively represent emotion strength, and by changing the emotion scalar, we can obtain different pitch, speech rate and tone patterns for different strength levels.

\vspace{-7pt}
%\section{SUMMARY}
\section{Summary}
\vspace{-3pt}

This paper proposes a controllable emotion speech synthesis approach based on emotion embedding space learned from references. In order to deliver the emotion more accurate and expressive with strength control, we modify the Prosody-Tacotron structure with two emotion classifiers to enhance the emotion-discriminative ability of the emotion embedding and the predicted mel-spectrum. Moreover, we adopt a style loss to measure the difference between the generated and reference mel-spectrum. During inference, the strength of the synthetic speech can be easily controlled by adjusting a scalar to the emotion embedding. Comparative experiments with other methods show that the synthetic speech of the proposed method is more accurate and expressive with less emotion confusions and the emotion strength control is more salient to subjective listeners. Samples can be found from \textcolor{blue}{https://silyfox.github.io/iscslp-98-demo/}.

\bibliographystyle{IEEEtran}

\bibliography{mybib}

% \begin{thebibliography}{9}
% \bibitem[1]{Davis80-COP}
%   S.\ B.\ Davis and P.\ Mermelstein,
%   ``Comparison of parametric representation for monosyllabic word recognition in continuously spoken sentences,''
%   \textit{IEEE Transactions on Acoustics, Speech and Signal Processing}, vol.~28, no.~4, pp.~357--366, 1980.
% \bibitem[2]{Rabiner89-ATO}
%   L.\ R.\ Rabiner,
%   ``A tutorial on hidden Markov models and selected applications in speech recognition,''
%   \textit{Proceedings of the IEEE}, vol.~77, no.~2, pp.~257-286, 1989.
% \bibitem[3]{Hastie09-TEO}
%   T.\ Hastie, R.\ Tibshirani, and J.\ Friedman,
%   \textit{The Elements of Statistical Learning -- Data Mining, Inference, and Prediction}.
%   New York: Springer, 2009.
% \bibitem[4]{YourName17-XXX}
%   F.\ Lastname1, F.\ Lastname2, and F.\ Lastname3,
%   ``Title of your INTERSPEECH 2020 publication,''
%   in \textit{Interspeech 2020 -- 20\textsuperscript{th} Annual Conference of the International Speech Communication Association, September 15-19, Graz, Austria, Proceedings, Proceedings}, 2020, pp.~100--104.
% \end{thebibliography}

\end{document}